\documentclass[a4paper,11pt]{article}
\usepackage{jheppub} 
\usepackage[normalem]{ulem}
\usepackage{xcolor}
\usepackage{colortbl}
\usepackage{multirow}
\usepackage{slashed}
\usepackage{dsfont}
\usepackage{caption}
\usepackage{subcaption}

\newcommand{\be}{\begin{eqnarray}}
\newcommand{\ee}{\end{eqnarray}}
\newcommand{\bea}{\begin{eqnarray}}
\newcommand{\eea}{\end{eqnarray}}
\newcommand{\nn}{\nonumber}

\newcommand{\beq}{\begin{equation}}
\newcommand{\eeq}{\end{equation}}

\newcommand{\five}{\mathbf{5}}
\newcommand{\three}{\mathbf{3}}
\newcommand{\two}{\mathbf{2}}

\newcommand{\MGUT}{M_{\rm GUT}}
\newcommand{\LQCD}{\Lambda_{\rm QCD}}
\newcommand{\MSUSY}{\widetilde M_S}
\newcommand{\aGUT}{\alpha_{\rm GUT}}

\definecolor{darkblue}{rgb}{0.2,0.2,0.9}
\definecolor{colorRTD}{rgb}{.2,.2,.7}

\definecolor{colorHD}{rgb}{.2,0.9,.0.9}

\title{A Cosmological Solution to the Doublet-Triplet Splitting Problem}

\author[1]{Csaba Csaki,}
\affiliation[1]{Department of Physics, LEPP, Cornell University, Ithaca, NY 14853, USA}
\author[2,3]{Raffaele Tito D'Agnolo,}
\affiliation[2]{Universit\'e Paris-Saclay, CEA, CNRS, Institut de Physique Th\'eorique, 91191, Gif-sur-Yvette, France}
\affiliation[3]{Laboratoire de Physique de l'\'Ecole Normale Sup\'erieure, ENS, Universit\'e PSL, CNRS, Sorbonne Universit\'e, Universit\'e Paris Cit\'e, F-75005 Paris, France}
\author[1,4]{Eric Kuflik,}
\affiliation[4]{Racah Institute of Physics, Hebrew University of Jerusalem, Jerusalem 91904, Israel}
\author[2]{Pablo Sesma}

\emailAdd{csaki@cornell.edu}
\emailAdd{raffaele-tito.dagnolo@ipht.fr}
\emailAdd{eric.kuflik@mail.huji.ac.il}
\emailAdd{pablo.sesma@ipht.fr}

\abstract{We propose a model that provides a simultaneous solution to the doublet-triplet splitting problem of grand unified theories, the electroweak hierarchy problem and the strong CP problem. The mechanism is based on the dynamics of two axion-like particles that would crunch the universe at the time of the QCD phase transition if triplets were light or had a VEV or if doublets were heavy or did not have a VEV. The only trace left at low energies are these two axion-like particles. They are weakly coupled to the Standard Model and could be detected at upcoming axion experiments or by a combination of neutron EDM measurements and the astrophysical detection of fuzzy dark matter.}

\begin{document}

\maketitle

\flushbottom

\section{Introduction}

The issue of naturalness has played a central role in particle physics over the past four decades. Traditional approaches attempt to explain the hierarchy between the weak scale and any high scale, such as the GUT or Planck scales, by introducing new physics around a TeV, which will modify the behavior of quantum effects once these news particles become accessible. 

However, no new particles at scales around $\sim 1-3$ TeV have been observed at the LHC, putting serious pressure on the most commonly considered TeV-scale physics scenarios, such as supersymmetric extensions of the Standard Model (SM)~\cite{Dimopoulos:1981zb, Nilles:1983ge, Haber:1984rc, Martin:1997ns} or composite Higgs models~\cite{Kaplan:1983fs, Kaplan:1983sm, Georgi:1984ef, Dugan:1984hq, Giudice:2007fh} and their variations~\cite{Arkani-Hamed:2001nha, Arkani-Hamed:2002sdy, Arkani-Hamed:2002ikv}. While some reasonable options for such models still remain (for example models of neutral naturalness~\cite{Chacko:2005pe, Burdman:2006tz, Craig:2015pha}), it is important to explore alternative approaches. 

The most promising such directions are referred to as models of cosmological naturalness and include relaxion-type models~\cite{Graham:2015cka}, where during inflation the Higgs potential is scanned via a rolling scalar, which is then stopped by a phase transition involving the VEV of the Higgs. Other interesting ideas are the so called crunching models~\cite{Bloch:2019bvc, Csaki:2020zqz, TitoDAgnolo:2021nhd, TitoDAgnolo:2021pjo}, where different patches of the Universe have different parameters for the Higgs potential, but some form of dynamics makes the patches with small or large Higgs VEVs dynamically unstable, leading to a rapid crunch. In this paper we will be focusing on such crunching models, which have previously been successfully applied to address the ordinary hierarchy problem~\cite{Csaki:2020zqz, TitoDAgnolo:2021nhd, TitoDAgnolo:2021pjo}, and even provide a new approach to the cosmological constant problem~\cite{Bloch:2019bvc}. One particularly interesting approach is called Sliding Naturalness~\cite{TitoDAgnolo:2021nhd, TitoDAgnolo:2021pjo}, where the interplay of two (axion-like) light scalars with the dynamics of the SM ensures that only Universes with Higgs VEVs and QCD $\theta$-angles qualitatively similar to ours survive for a sufficiently long time to allow structure formation.

There are many reasons to expect that grand unification~\cite{Georgi:1974sy, Fritzsch:1974nn, Dimopoulos:1981zb, PhysRevD.10.275} will be an important ingredient in whatever form of UV theory is realized in Nature. Hypercharge is not asymptotically free, and without unification at some very high scale it is expected to hit a Landau pole. Charge quantization and the observed SM quantum numbers are also calling for Grand Unified Theories (GUTs). Additionally, GUTs provide a natural environment to incorporate magnetic monopoles, which could be an alternative explanation for charge quantization. However the simplest GUTs (supersymmetric or not) usually have their own naturalness problem in addition to the usual Higgs hierarchy problem. The issue is how to successfully split the Higgs doublet from its GUT partners, in particular any additional triplets, which would generically destroy unification and also may give rise to proton decay. This goes under the name of doublet-triplet (D-T) splitting problem, and usually requires some additional intricate mechanism to be solved, which will usually make the GUT models overly complicated. 

In this paper we will explore how the Sliding Naturalness model~\cite{TitoDAgnolo:2021nhd, TitoDAgnolo:2021pjo} can be embedded in the simplest GUT models, and in particular how the D-T splitting problem is affected by the additional axion-like singlets. As in other crunching models, we imagine that a Multiverse scans some of the dimensionful parameters in the Higgs doublet/triplet potentials. We will show that the same crunching mechanism that selects universes with light doublet Higgses getting VEVs will also  ensure that the Higgs triplets are heavy and do not have VEVs. For this, we will be examining the modification of the SM dynamics in hypothetical universes where the Higgs triplets would be light and/or get VEVs, or where the Higgs doublets are heavy and/or VEVless. We will show that only the universes qualitatively similar to ours (light doublet Higgses with VEVs and heavy triplets without VEV) will be long lived and all others are expected to quickly crunch. Additionally, if the QCD $\theta$-angle scans across the Multiverse, universes with a large $\bar \theta$ will also crunch.
To prove that the crunching mechanism works, we will account for the details of the induced Higgs-dependent singlets potentials, considering both the confining dynamics and small instantons. 
While we will be using a supersymmetric example to simplify the discussion and ensure a precise form of unification, SUSY will not be playing a major role in solving the Higgs hierarchy problem. 

The paper is organized as follows: in Section~\ref{sec:problem} we review the D-T splitting problem with emphasis on what happens in a Multiverse. In Section~\ref{sec:model} we present our supersymmetric GUT model and list the qualitatively different types of universes in our Multiverse. In Sections~\ref{sec:basic} and \ref{sec:sliding} we show how our Sliding Naturalness model embedded in a supersymmetric GUT solves simultaneously the doublet-triplet splitting problem, the hierarchy problem and the strong CP problem. In Section~\ref{sec:potential} we present a detailed calculation of the axion-like potentials in our Multiverse. This calculation is at the core of the solutions of the three problems in the previous Sections. We conclude and comment on the phenomenology of the model in Section~\ref{sec:conclusion}.

\section{The Doublet-Triplet Splitting Problem}
\label{sec:problem}
The only dimensionful scale appearing in the SM Lagrangian is the mass $\mu$ in the Higgs potential
\be
V=-\mu^2 |H|^2 + \lambda |H|^4 , \label{muterm}
\ee
where $H$ is the electroweak Higgs doublet. One may expect this parameter 
to be of order the fundamental scale of the theory, such as the Planck scale or GUT scale, but we observe the Higgs doublet to be much lighter. 
    In grand unified theories there is an additional problem. 
 In the minimal $SU(5)$ model the Higgs fields can be embedded in a $\five$ multiplet, but this implies SU(5) partners of the Higgs:
 \be
\five_H \to (\three, 1)_{-1/3} \oplus (1, \two)_{1/2}  = T  \oplus  H\,.
\ee
While the Higgs doublet needs to be light, the \emph{Higgs triplet} ($T$) must be much heavier than the weak scale. Integrating out the triplet at tree-level generates the dimension six operator\footnote{The two Yukawa couplings $Y_{uT}, Y_{dT}$ are associated to the terms in the Lagrangian
\be \mathcal{L}\supset -\frac{1}{2}Y_{uT} QQ T + Y_{dT} QL T^\dagger\, .\ee At the GUT scale they are equal to the SM up and down Yukawa couplings, respectively.
} $(Y_{uT} Y_{dT}/m_T^2) QQQL$, which can induce rapid proton decay for $m_T \simeq m_W$. Additionally, heavy triplets are needed for consistent unification.
However, the $SU(5)$ symmetry dictates that the Higgs doublet and triplet are degenerate. This is the doublet-triplet splitting problem. We can make it more concrete by looking at the Lagrangian of a simple model.

In minimal  $SU(5)$ the masses for the Higgs multiplets arise from the scalar potential terms 
\be
V= -\left({\mu_5^2} + \alpha {\rm Tr}[\Sigma^2] \right) \five_H^\dagger\five_H + \lambda (\five_H^\dagger\five_H )^2+ \beta \five_H^\dagger\Sigma^2 \five_H  +\delta \mu_\Sigma \five_H^\dagger\Sigma \five_H \,,
\label{eq:SU51}
\ee
where $\Sigma$ is a scalar that transforms as a $\mathbf{24}$ of $SU(5)$ and is responsible for spontaneously breaking $SU(5)$ down to the SM gauge groups with the VEV
\be
\langle \Sigma \rangle = v_{\Sigma} \; {\rm diag}\left(-2,-2,-2, 3,3\right), \label{eq:vSigma}
\ee
where $v_{\Sigma}\sim \MGUT$.
After $SU(5)$ breaking the potential becomes 
\be
V &=& \left[ -{\mu_5^2} + v_\Sigma^2\left(-30 \alpha+ 4\beta - \frac{2 \delta \mu_\Sigma}{v_\Sigma}\right)\right] H^\dagger H +   \left[ -{\mu_5^2} + v_\Sigma^2\left(-30 \alpha+ 9\beta +\frac{3 \delta \mu_\Sigma}{v_\Sigma}\right)\right] T^\dagger T\nn\\  &+&\lambda\left (T^\dagger T +H^\dagger H \right)^2\,.
\label{eq:su5mu}
\ee
To get a light electroweak scale while keeping the triplets near the GUT scale, at tree level the parameters must be chosen so that 
\be
m_H^2 &= &   {\mu_5^2} - v_\Sigma^2\left(-30 \alpha+ 4\beta - \frac{2 \delta \mu_\Sigma}{v_\Sigma}\right)= \mathcal{O} (m_h^2) \label{eq:tuning1}\,,\\
 m_T^2 &=& {\mu_5^2} - v_\Sigma^2\left(-30 \alpha+ 9\beta +\frac{3 \delta \mu_\Sigma}{v_\Sigma}\right)  = \mathcal{O} (\MGUT^2).  \label{eq:tuning2}
\ee

The exact same problem arises also in supersymmetric models of grand unification. If we consider again a minimal $SU(5)$ model, the Higgs fields $H_u$ and $H_d$ can be found in the $\five$ and $\bar \five$  multiplets, respectively, but these imply SU(5) partners of the Higgses:
 \be
H_\five &\to& (\three, 1)_{1/3} \oplus (1, \two)_{1/2}  = T_u  \oplus  H_u\nn \\
H_{\bar \five} &\to& (\bar \three, 1)_{1/3} \oplus (1,\bar\two)_{-1/2}  = T_d  \oplus  H_d\,.
\ee
As in the non-supersymmetric case, the Higgs doublets need to be light for correct electroweak symmetry breaking, the Higgs triplets ($T_u,T_d$) must be close to the GUT scale to slow down proton decay and for successful unification in the MSSM. Proton decay is mediated by the usual dimension six operator $(Y_{uT} Y_{dT}/m_T^2) QQQL$ and by a dimension five operator generated when integrating out the fermionic partners of the triplets~\cite{Murayama:2001ur}.

In the supersymmetric case, the masses for the Higgs multiplets come from the superpotential terms
\be
\mathcal{W}\supset \lambda H_{\bar 5} \Sigma H_5 + \mu_5 H_{\bar 5} H_5\, . \label{eq:SU51}
\ee
After $SU(5)$ breaking the superpotential becomes 
\be
\mathcal{W}\supset \left(\mu_5 -2\lambda v_\Sigma\right)T_d T_u +\left(\mu_5+3\lambda v_\Sigma\right)H_u H_d\, , \label{eq:su5mu}
\ee
where we normalize the GUT-breaking VEV of $\Sigma$ as in Eq.~\eqref{eq:vSigma}. 
To get a light electroweak scale while keeping the triplets at the GUT scale, the parameters must be chosen so that 
\be
\mu \equiv  \mu_5+3\lambda v_\Sigma = \mathcal{O} (m_h)\,, \qquad \qquad \mu_5 -2\lambda v_\Sigma= \mathcal{O} (\MGUT).  \label{eq:tuningSUSY}
\ee
Although this is a radiatively stable choice in supersymmetry, it is still a large tuning between $\mu_5$ and $ \lambda v_\Sigma$. Thus we found the same tuning required in the non-supersymmetric model. A similar problem arises in unified models with different gauge groups.

Several approaches to the doublet-triplet splitting problem exist in the literature. These include the Missing Partner Mechanism~\cite{Masiero:1982fe, Grinstein:1982um, Altarelli:2008bg},
extra dimensional GUT's~\cite{Kakizaki:2001en, Hall:2001xr, Faraggi:2001ry, Maru:2001ch, Shafi:2001pz},
the Missing VEV (or Dimopoulos-Wilczek) mechanism~\cite{Dimopoulos:1981zu, Srednicki:1982aj} in $SO(10)$,
the pseudo-Goldstone boson mechanism~\cite{Berezhiani:1995sb},
and the sliding singlet mechanism~\cite{Witten:1981kv, Nanopoulos:1982wk, Dimopoulos:1982af, Ibanez:1982fr, Nemeschansky:1983gw}.
While each of these contains a very clever idea, they all suffer from some type of drawback - either complicated group theory, some hidden tuning or a dynamical instability.

In the following we introduce a mechanism that naturally explains the large hierarchy between the Higgs doublets and the triplets. 
While the mechanism can work for general non-SUSY GUTs, to simplify our discussion and to most easily ensure the unification of couplings we will be assuming a SUSY theory throughout the paper, and explain how our mechanism works in SUSY models.

\subsection{The Doublet-Triplet Splitting Problem in the Multiverse}
Before turning to our model, it is worthwhile  spending a few words on the problem in the Multiverse. In the traditional setup, where the smallness of the EW scale is explained by a symmetry, the D-T splitting problem is an additional tuning arising in minimal GUTs. In any minimal GUT, the Higgs mass receives a $\mathcal{O}(\MGUT)$ contribution at tree-level from the VEV of $\Sigma$, as shown in Eq.~\eqref{eq:tuningSUSY} for a SUSY scenario, no matter where the (super)symmetry-breaking scale is. Unfortunately $\MGUT \gg m_W$ due to constraints coming mainly from proton decay and unification, so making $m_h$ small requires a tuning even if the SUSY-breaking scale is at the weak scale. 

In a Multiverse the situation is slightly different. If we imagine that the GUT scale is a fundamental scale, fixed throughout the Multiverse, and the smallness of the EW scale is due to a tuning, this same tuning can also solve the doublet-triplet splitting problem. All that is needed is to tune the mass of the SM-like Higgs to be small and this needs to be done only once. There aren't two separate tunings for the hierarchy and D-T problems. Once this tuning is realized the triplets remain naturally at the GUT scale. 

The mechanism presented here explains why we observe this tuning, rather than living in a universe where all scalars are at the fundamental scale $\MGUT$. Additionally, it also explains why, out of all possible universes where one scalar is light, we observe one where a doublet is light and has a VEV, rather than a universe with a light triplet with or without a VEV or a universe with a light doublet without a VEV. 

\section{The GUT Model}\label{sec:model}

In this work we envision that there is a Multiverse which scans the tuning in Eq.~\eqref{eq:tuningSUSY}. Technically, everything in this paper is done assuming that  $\mu_5$ in Eq.~\eqref{eq:SU51} scans and no other parameter varies from universe to universe. That is, $\lambda v_\Sigma \sim \MGUT$ and the SUSY breaking scale $\MSUSY$ are kept fixed, and so are the dimensionless couplings (gauge, Yukawa and quartics) at the GUT scale. A way to scan dimensionful parameters without scanning appreciably dimensionless ones, was realized in the ``friendly lanscape" framework of~\cite{Arkani-Hamed:2005zuc}. We imagine that the SUSY breaking scale and the dimensionful parameters in the $\Sigma$ potential do not scan because they are generated via dimensional transmutation. The only dimensionless parameter that we allow to vary at $\mathcal{O}(1)$ from universe to universe is the QCD $\theta$-angle, to show that the mechanism can also account for its small observed value. However, this assumption is not necessary for the mechanism to work, we could relax $\bar \theta$ to zero throughout the Multiverse with an axion and the solutions to D-T and hierarchy problems would be unaffected.

Furthermore, we take the unified gauge group and particle content at $\MGUT$  to be the same in all universes. 
This is not crucial for our discussion, but allows to do explicit calculations and makes it more concrete.

To realize gauge coupling unification
we take all superpartners to be at the same scale, $\MSUSY \simeq 10^6$~GeV. This point gives good unification with $\MGUT\simeq 1.3\times10^{16}$~GeV and $\aGUT^{-1}\simeq 25.3$. Additionally, choosing such a high value for the SUSY breaking scale emphasizes that in these models the hierarchy problem is not solved by SUSY - it is the Sliding Naturalness mechanism that is responsible for the solution. SUSY is merely added for the sake of precise unification and to simplify the structure of the Higgs potential. There would be no obstacle to choosing a smaller value of $\MSUSY$ all the way down to the TeV scale (as long as the experimental bounds on $\MSUSY$ are satisfied). 

Between $\MGUT$ and $\MSUSY$ we have a supersymmetric gauge theory described by the following superpotential
\be
\mathcal{W}&=& \lambda H_{\bar 5} \Sigma H_5 + \mu_5 H_{\bar 5} H_5 + \frac{Y_{10}}{8} \Phi_{10}\Phi_{10} H_5 + Y_{5} \Phi_{\bar 5}\Phi_{10} H_{\bar 5} \nn \\
&+& \frac{M_\Sigma}{2}{\rm Tr}[\Sigma^2]+\frac{\lambda_\Sigma}{3}{\rm Tr}[\Sigma^3]\, , \label{eq:SUSYSU5}
\ee
where $\Phi_{10}$ and $\Phi_{\bar 5}$ are the superfileds containing the SM fermions in their usual $SU(5)$ embedding~\cite{Georgi:1974sy, Dimopoulos:1981zb}. $Y_{10}$ and $Y_5$ become the SM Yukawa couplings, $Y_u$ and $Y_{d,\ell}$ after running to the weak scale.
The global minimum of the scalar potential is at~\cite{Dimopoulos:1981zb}
\be
\langle H_5 \rangle = \langle H_{\bar 5} \rangle = \langle \Phi_{10} \rangle = \langle \Phi_{\bar 5} \rangle = 0\, . \label{eq:vscalars}
\ee
There are three possibilities for the VEV of $\Sigma$~\cite{Dimopoulos:1981zb} and we assume that the theory selects dynamically the one that breaks $SU(5)$ to the SM, given in Eq.~\eqref{eq:vSigma}.

It is easy to conclude from Eq.~\eqref{eq:tuningSUSY} that in this theory the doublets and triplets cannot be simultaneously light. In a universe where the doublets are tuned to be light, at the SUSY breaking scale we have the MSSM Higgs sector 
\be
V_{H_{u,d}} &=& m_U^2 |H_u|^2 + m_D^2 |H_d|^2 - B\mu (H_u H_d +{\rm h.c.}) + \frac{g^2}{2} |H_u H_d|^2 \nn \\
&+&\frac{g^2+g^{\prime 2}}{8}\left(|H_u|^2-|H_d|^2\right)^2\, ,\label{eq:MSSM} \\
m_{U,D}^2&=&|\mu_5+2 \lambda v_\Sigma|^2+m_{H_{u,d}}^2\, ,\label{eq:mUD}
\ee
where $m_{H_{u,d}}^2$ are the usual soft SUSY breaking masses.
Without loss of generality, we can set  the VEVs of the charged components of $H_{u,d}$ to zero, using the $SU(2)_L$ gauge symmetry, and focus on the neutral components of the two doublets. 

At this point we make one more model-building choice and take $B\mu$ to be slightly smaller than the SUSY breaking scale, $B\mu = \epsilon \MSUSY^2$, with $1/50 \lesssim \epsilon \lesssim 1$. This is technically natural as $B\mu$ breaks a PQ symmetry of the Higgs sector. We further assume that this little hierarchy between $\MSUSY$ and $B\mu$ is not scanned in the Multiverse, as it arises from a dimensionless PQ-breaking coupling.

\begin{figure}[!t]
\begin{center}
\includegraphics[width=0.95\textwidth]{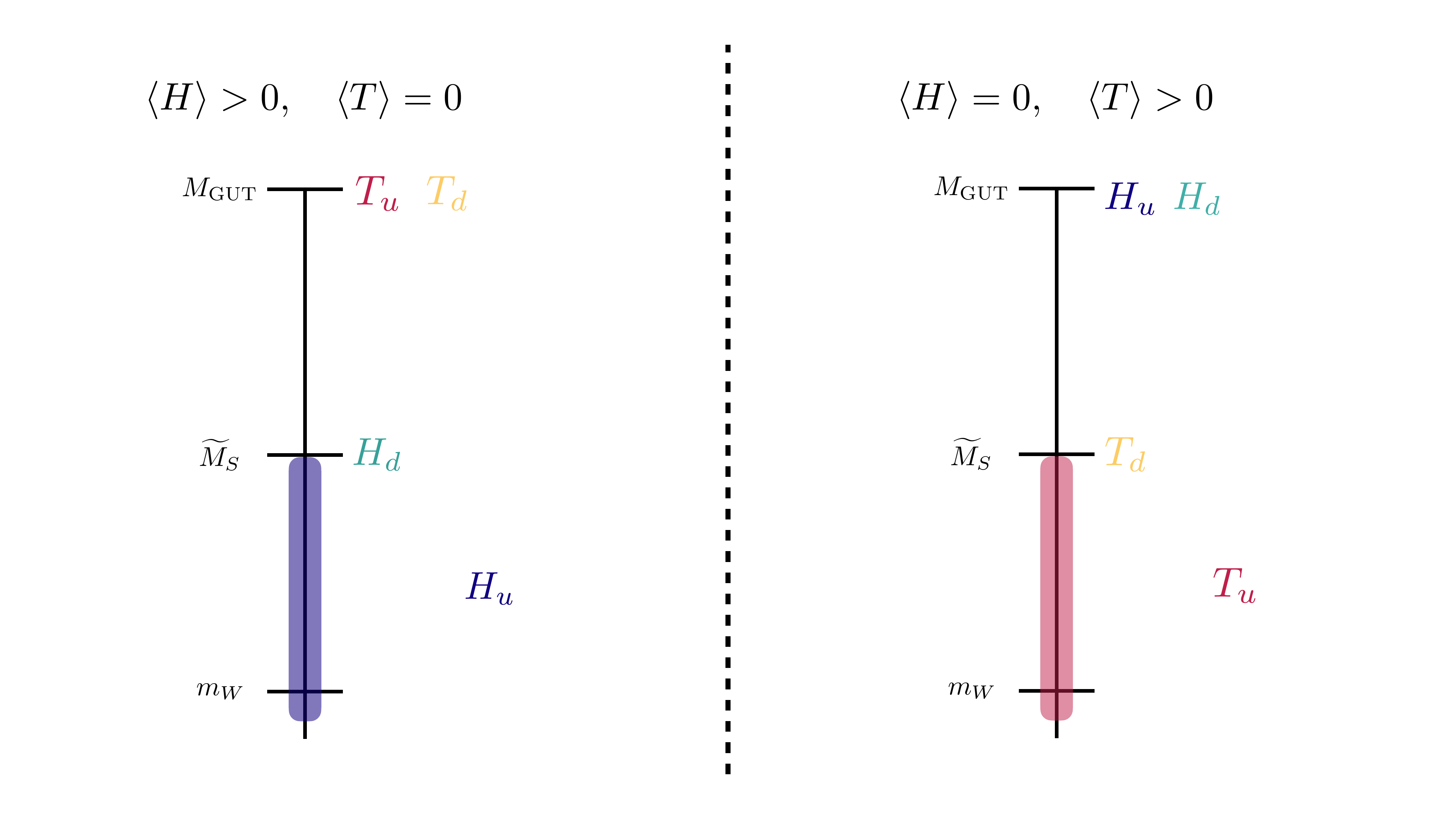}
\caption{Universes with VEVs that can exist in our Multiverse. Either $H_{u,d}$ or $T_{u,d}$ must be at the GUT scale. Scalar VEVs can only exist at or below the SUSY-breaking scale $\MSUSY$. If one pair of scalars (either $H_{u,d}$ or $T_{u,d}$) is tuned to be light, one of them must be around $\MSUSY$ (here for concreteness $H_d$ or $T_d$), the other can be lighter. Other universes either do not exist, given our model in Section~\ref{sec:model}, or are more tuned. The shaded area indicates that the scalar with the same color can be at any scale within that area.}
\label{fig:uni1}
\end{center}
\end{figure}

In universes like our own, with an electroweak scale, 
\be
m_Z^2 = (m_U^2+m_D^2)\left(\frac{|m_U^2-m_D^2|}{\sqrt{(m_U^2+m_D^2)^2-4 B\mu^2}}-1\right)\, ,
\ee
which is parametrically smaller than $\MSUSY^2$, we always have $B\mu^2 \simeq m_U^2 m_D^2$. The other option for making $m_Z^2\ll \MSUSY^2$, i.e. $m_U^2+m_D^2 \ll \MSUSY^2$, cannot describe our universe, since $m_U^2+m_D^2\lesssim B\mu$ gives rise to an unstable potential along its D-flat direction, and we assumed that $B\mu\simeq \epsilon \MSUSY^2 \gg 100$~GeV, is fixed throughout the Multiverse. 

Making\footnote{In our Multiverse this tuning is realized by scanning $\mu_5$ in Eq.~\eqref{eq:mUD}} $m_U^2 m_D^2 \simeq B\mu^2$, while keeping $m_U^2+m_D^2=\mathcal{O}(\MSUSY^2)$, has two physical consequences: 1) All Higgses are at the SUSY breaking scale, $m_A^2 \simeq m_{H^\pm}^2 \simeq m_{H^0}^2 \simeq m_U^2 + m_D^2 = \mathcal{O}(\MSUSY^2)$, except for one CP-even Higgs at the EW scale $m_{h^0}^2\simeq m_Z^2$, 2) There is a hierarchy between the VEVs of the two doublets, 
\be
\sin(2\beta)=\frac{2\langle H_d^0\rangle \langle H_u^0\rangle}{\langle H_u^0\rangle^2+\langle H_d^0\rangle^2}=\frac{B\mu}{m_U^2+m_D^2}\simeq \frac{m_U m_D}{m_U^2+m_D^2} = \mathcal{O}(\epsilon)\, , \label{eq:vevs}
\ee
and the lightest doublet dominates EW symmetry breaking. If we take for instance $m_U^2 \ll m_D^2$, then
\be
\frac{\langle H_d^0\rangle}{\langle H_u^0\rangle} \simeq \frac{B\mu}{m_D^2} \simeq \epsilon\, .
\ee
This is one of the standard decoupling limits of the 2HDM~\cite{Gunion:2002zf}, considered also in~\cite{Arkani-Hamed:2004ymt}.
We will often talk about universes where $\langle H \rangle > 0$ and $\langle T \rangle = 0$ to refer to universes like the one that we have just described with $T_{u,d}$ at the GUT scale, $H_d$ (or $H_u$) at $\widetilde M_S$, and $H_u$ (or $H_d$) at or below $\MSUSY$.

\begin{figure}[!t]
\begin{center}
\includegraphics[width=0.95\textwidth]{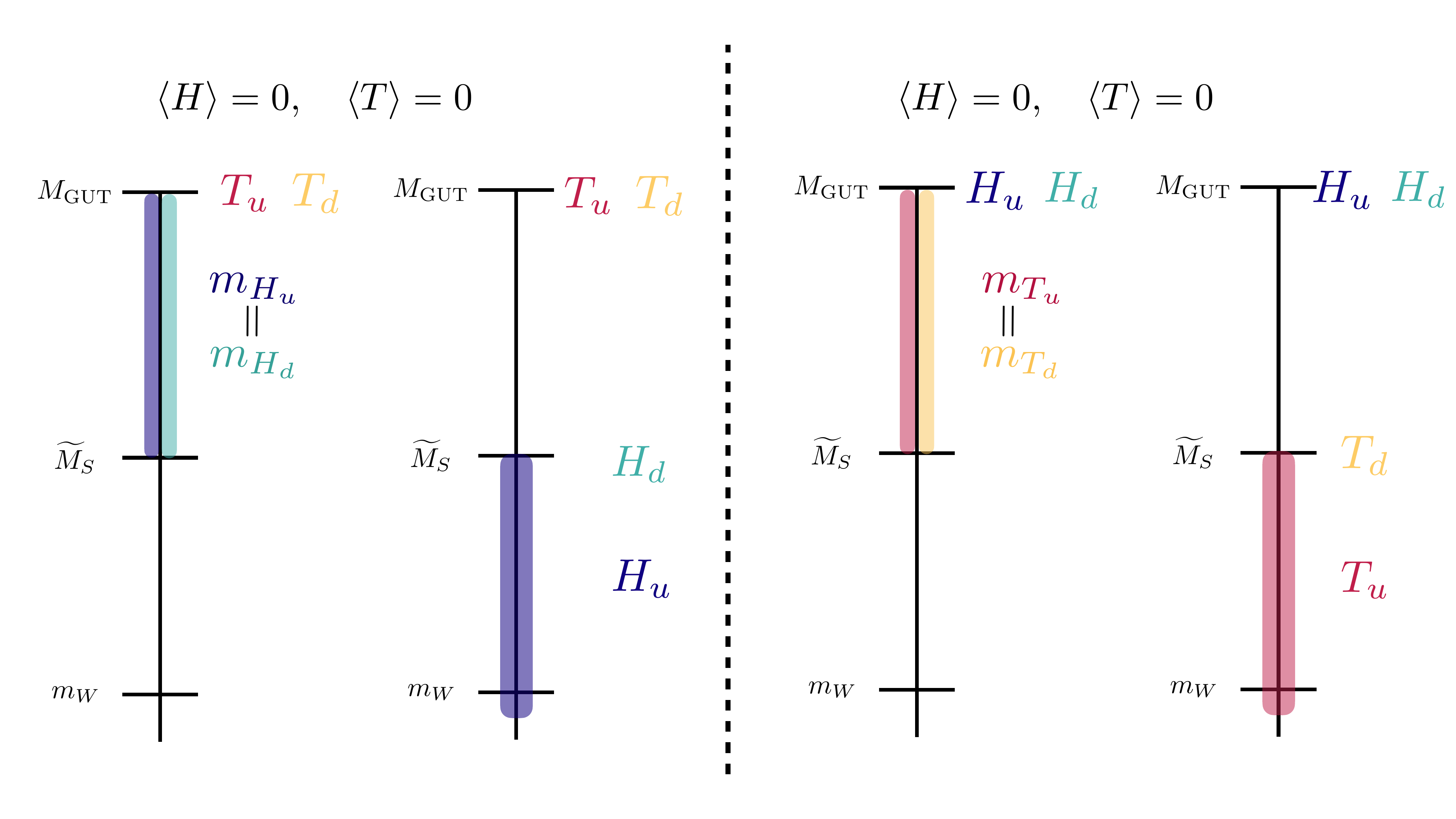}
\caption{Universes without VEVs that can exist in our Multiverse. Either $H_{u,d}$ or $T_{u,d}$ must be at the GUT scale.  If one pair of scalars (either $H_{u,d}$ or $T_{u,d}$) is tuned to be light, they have the same mass if they are tuned above $\MSUSY$ (left subfigure of each panel). If one is tuned parametrically below $\MSUSY$ the other must be around $\MSUSY$ (right subfigure of each panel). The shaded area indicates that the scalar with the same color can be at any scale within that area.}
\label{fig:uni2}
\end{center}
\end{figure}

Summarizing, we are imagining a Multiverse that scans $\mu_5$ in Eq.~\eqref{eq:SUSYSU5}, and where $\MGUT^2 \gg \MSUSY^2 \gtrsim B\mu$ are kept fixed. We make the same assumptions on the SUSY breaking parameters in the triplets sector (i.e. the $B\mu$-term for the triplets is also smaller than $\MSUSY$). In this Multiverse doublets and triplets cannot be simultaneously light and cannot simultaneously have a VEV. This is due to the structure of the minimal $SU(5)$ superpotential. As recalled in Eq.~\eqref{eq:vscalars} in the supersymmetric theory only $\Sigma$ can have a VEV, the other scalars can have one only if supersymmetry is broken. Therefore, every time that doublets or triplets have a VEV, this occurs at or below the SUSY breaking scale. Additionally, since we are scanning only $\mu_5$, there is no way to have both doublets and triples light as shown in Eq.~\eqref{eq:tuningSUSY}. Therefore we have four qualitatively different types of universes summarized in Fig.s~\ref{fig:uni1} and~\ref{fig:uni2}.

In the left panel of Fig.~\ref{fig:uni1} we have universes where the Higgs doublets have a VEV and the triplets do not. In these universe the triplets $T_{u,d}$ are both at the GUT scale, one doublet is at $\MSUSY$ and a second doublet dominates EW symmetry breaking if it is below $\MSUSY$. We can also have both doublets at $\MSUSY$, possibly with comparable VEVs.

In the right panel of Fig.~\ref{fig:uni1} we have universes where the role of the triplets and doublets is inverted compared to the left panel that we have just described. Here $SU(3)_c$ is broken to $SU(2)_c$, because, when the triplets have a VEV, we can set to zero the VEVs of two components of the triplet, using the $SU(3)_c$ symmetry, without loss of generality.
 
The last two kinds of universes are described in Fig.~\ref{fig:uni2}. If neither triplets nor doublets have a VEV, we still have one pair at the GUT scale, either $T_{u,d}$ (left panel) or $H_{u,d}$ (right panel). The other pair can have any mass, but if this mass is above $\MSUSY$, it is the same for both up- and down-type scalars, because it comes from the effective $\mu$-term in the superpotential in Eq.~\eqref{eq:su5mu}. Instead, if one scalar is tuned below $\MSUSY$, the other has to be at $\MSUSY$, as discussed above Eq.~\eqref{eq:vevs}.

In the next Section we show how to solve dynamically the doublet-triplet splitting problem in this Multiverse (and the electroweak hierarchy and strong CP problems). The vast majority of universes have both $H_{u,d}$ and $T_{u,d}$ at the GUT scale and we have to explain the tuning that makes one doublet light in our universe, without accidentally selecting also unwanted universes in Fig.s~\ref{fig:uni1} and~\ref{fig:uni2}.

\section{Basic Idea and Summary of Results}\label{sec:basic}

Here we ask if the tuning described in the previous Section can be the result of cosmological dynamics, similar to cosmological solutions to the SM hierarchy problem. 
In fact,  solutions that rely on the Higgs VEV to trigger a cosmological event, naturally lend themselves as solutions to the doublet-triplet splitting problem. This is because the same dynamics can be made sensitive to not only the Higgs doublet mass and VEV, but also to the same properties of the Higgs triplet. 

\begin{figure}[!t]
\begin{center}
\includegraphics[width=\textwidth]{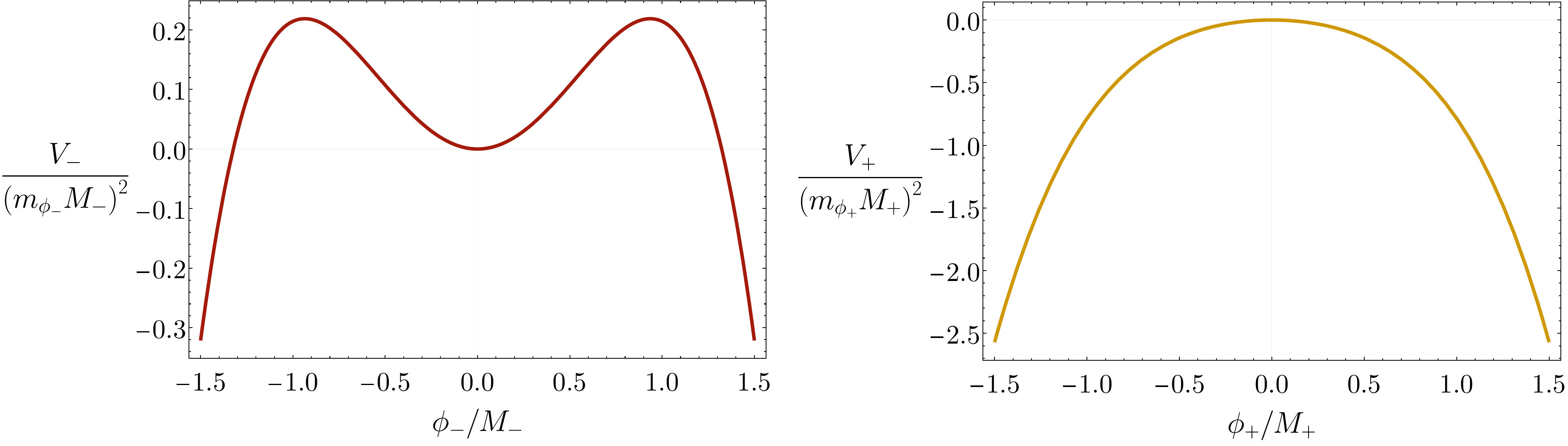}
\caption{$\phi_\pm$ potentials, excluding their couplings to the SM, around the locations of their local minima (the minimum of $\phi_+$ is generated only in our universe). $\phi_\pm$ roll down their potential and crunch all universes that do not have a light doublet with $\langle H \rangle \simeq 174$~GeV, a heavy triplet close to the GUT scale and a small $\bar \theta < 10^{-10}$.}
\label{fig:pot}
\end{center}
\end{figure}

There has been much work recently on early universe dynamics that explains the SM hierarchy by selecting our observed vacuum~\cite{Agrawal:1997gf, Dvali:2003br, Dvali:2004tma, Arkani-Hamed:2004ymt, Graham:2015cka, Espinosa:2015eda, Arvanitaki:2016xds, Geller:2018xvz,  Cheung:2018xnu, Giudice:2019iwl, Arkani-Hamed:2020yna, Strumia:2020bdy, Csaki:2020zqz, TitoDAgnolo:2021nhd, TitoDAgnolo:2021pjo, Giudice:2021viw, Khoury:2021zao, Chatrchyan:2022pcb, Trifinopoulos:2022tfx, Csaki:2022zbc, Matsedonskyi:2023tca, Hook:2023yzd, Chattopadhyay:2024rha}.   
We focus on the \emph{Sliding Naturalness}~\cite{TitoDAgnolo:2021nhd, TitoDAgnolo:2021pjo} explanation of the electroweak scale and neutron EDM.
The basic idea is summarized here, while the details are fleshed out in Section~\ref{sec:sliding}.  
We will assume that the low energy theory contains two new light scalars $\phi_\pm$, in addition to the matter fields considered in the previous Section. These scalars, near a local minimum ($\phi_-$) or maximum ($\phi_+)$ have the typical quadratic potential 
\be
V_\pm=\mp\frac{m_{\phi_\pm}^2}{2}\phi_\pm^2-\frac{m_{\phi_\pm}^2}{4 M_{\pm}^2}\phi_\pm^4+... \label{eq:pot1}\,,
\ee
depicted in Fig.~\ref{fig:pot}. At larger values of $\phi_\pm$, the $V_\pm$ potentials are UV-completed to stable potentials, as discussed in~\cite{TitoDAgnolo:2021nhd, TitoDAgnolo:2021pjo}.

The two scalars also have a Higgs-dependent potential whose dynamical origin is the axion-like interaction
\be
V_{H\phi} =- \frac{\alpha_5}{8\pi}\left(\frac{\phi_+}{F_+}+ \frac{\phi_-}{F_-}+\bar \theta \right){\rm Tr}[F_5\widetilde F_5]\, , \label{eq:Alike}
\ee 
where $F_5$ is the full $SU(5)$ field strength and $F_\pm$ the axion-like couplings of $\phi_\pm$.  
At low energies, the gauge theory dynamics generates an additional potential for $\phi_\pm$. We take $M_\pm \ll F_\pm$, which is technically natural and necessary for the mechanism to work, as explained in the next Section. Physically $M_-$ is roughly the distance between the two maxima in $V_-$ and $M_+$ the scale over which $V_+$ is approximately flat, as shown in Fig.~\ref{fig:pot}. In the $M_\pm \ll F_\pm$ limit, the potential in the region $|\phi_\pm|\lesssim M_\pm$ reads
\be
V_{H\phi} \simeq \frac{\Lambda^4}{2} \left(\frac{\phi_+}{F_+} + \frac{\phi_-}{F_-}+\bar\theta\right)^2+\cdots
\ee
As we show in the next Section, the total potential ($V_\pm+V_{H\phi}$) has stable local minima for both $\phi_\pm$ only when\footnote{We are imagining that the $\phi_-^2$ term from $V_{H\phi}$ can be neglected when $|\phi_-|\lesssim M_-$, as discussed in the next Section.}
\beq
\Lambda_{\rm min}\lesssim \Lambda \lesssim \Lambda_{\rm max}, \quad \bar\theta \lesssim \theta_{\rm max}\, . \label{eq:correctlambda}
\eeq
Generic values of $\Lambda$ and $\theta$ give unstable directions in the potential, leading to those universes/patches to collapse due to the large negative potential energy. Only universes with specific values of $\Lambda$ and $\theta$ survive for cosmologically long times and can give rise to galaxies. Observing our universe is a measurement of $\Lambda_{\rm min, max} \simeq \LQCD \simeq 0.1$~GeV and a measurement of $\theta_{\rm max} \simeq 10^{-10}$ that fix the parameters in the $\phi_\pm$ potentials. This is conceptually the same as measuring $\LQCD$ at low energy. Then fixing the theory up to $\Lambda_{\rm UV}$ will also fix the value of $\alpha_s(\Lambda_{\rm UV})$. In this scenario the weak scale and $\theta$ are our low energy parameters and the $\phi_\pm$ potentials are the UV dynamics that generates them.

If $\mu_5$ is scanned then  $\Lambda= \Lambda(\mu_5)$ varies in each patch, and the survival of the universe depends on the doublet-triplet mass splittings and VEVs. As discussed in the previous Section, we can divide the 4 types of universes in the previous Section into 3 qualitatively different categories: 
\begin{enumerate}
	\item $\left<H \right> >0 $ and $\left<T \right> =0 $, left panel of Fig.~\ref{fig:uni1}.
	\item $\left<H \right> =0 $ and $\left<T \right> >0 $, right panel of Fig.~\ref{fig:uni1}.
	\item $\left<H \right> =0 $ and $\left<T \right> =0 $, Fig.~\ref{fig:uni2}.
\end{enumerate}
We showed in the previous Section that when EW symmetry is broken both $H_{u,d}$ get a VEV, with the lightest doublet having the parametrically larger VEV. Here and in the following we use the shorthand notation $\left<H \right> >0$, $\left<T \right> =0$ to indicate universes such as those in the left panel of Fig.~\ref{fig:uni1}, where both doublets get a VEV, with one that can be parametrically larger than the other. The same is meant for triplets when we write $\left<H \right> =0$, $\left<T \right> >0$.

We find that in universes qualitatively different from our own one cannot obtain a scale $\Lambda$, for the $\phi_\pm$ potentials, comparable to  $\Lambda_{\rm SM}$, the corresponding scale in our universe,  by scanning $\mu_5$. Eq.~\eqref{eq:correctlambda} is satisfied only in our universe, where $\left<H \right> >0 $, $\left<T \right> =0 $, $0 < m_Z \ll \MSUSY,\MGUT $ and $m_{T_{u,d}} \simeq \MGUT $, and hence only universes qualitatively similar to ours are stable on cosmological timescales.  

\begin{figure}[!t]
\begin{center}
\includegraphics[width=\textwidth]{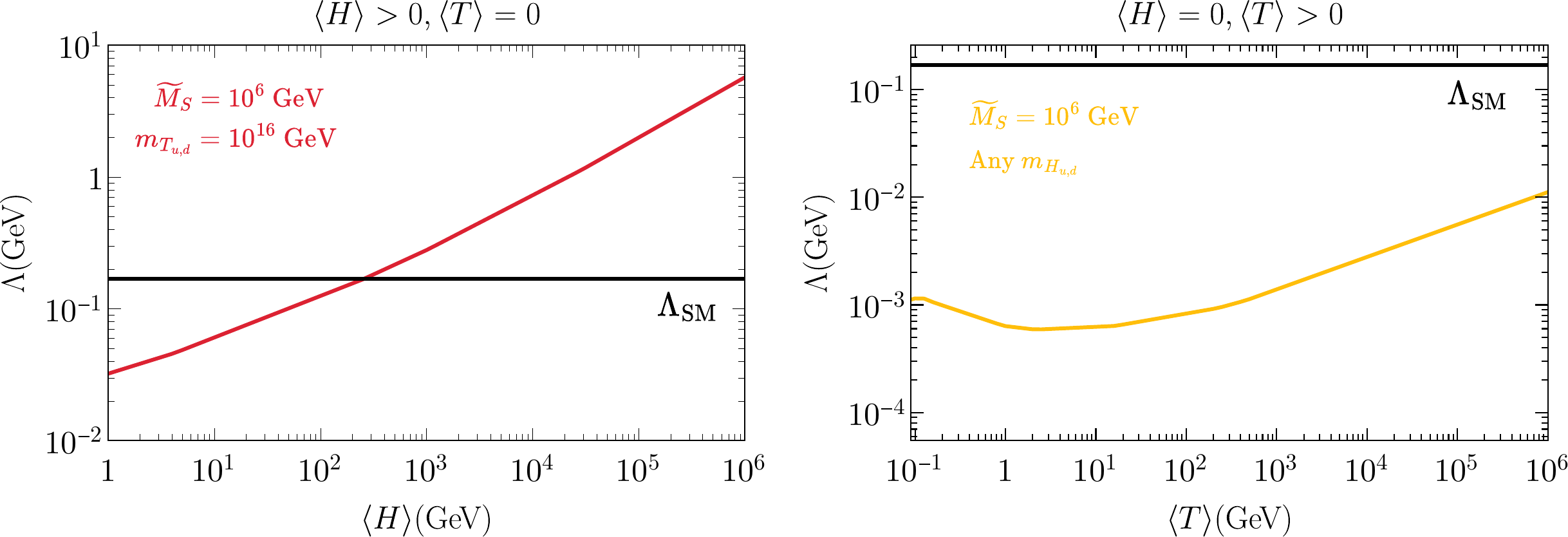}
\caption{Scale of the $\phi_\pm$ potentials $\Lambda$ in universes with VEVs, compared to the scale in our universe, $\Lambda_{\rm SM}$. The structure of these universes is summarized in Fig.~\ref{fig:uni1}. Universes with $\Lambda$ very different from $\Lambda_{\rm SM}$ crunch shortly after the QCD phase transition.}
\label{fig:LQCD1}
\end{center}
\end{figure}

\begin{figure}[!t]
\begin{center}
\includegraphics[width=\textwidth]{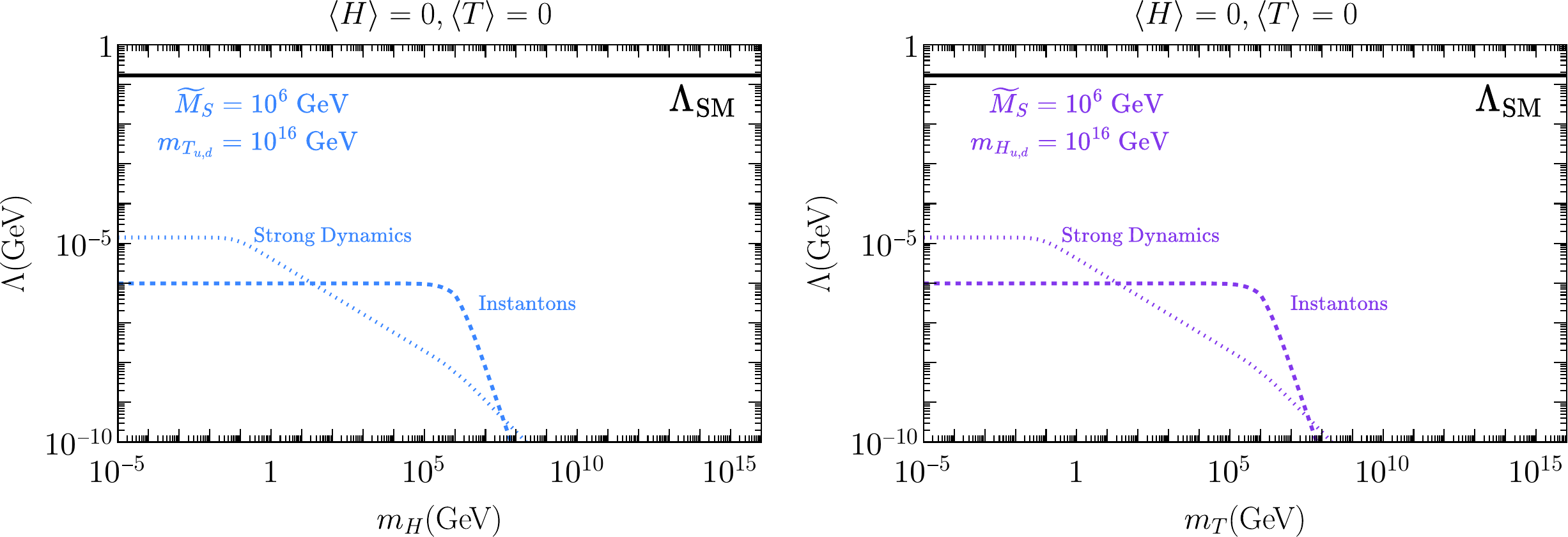}
\caption{Scale of the $\phi_\pm$ potentials $\Lambda$ in universes without VEVs compared to the one in our universe, $\Lambda_{\rm SM}$. The structure of these universes is summarized in Fig.~\ref{fig:uni2}. Universes with $\Lambda$ very different from $\Lambda_{\rm SM}$ crunch shortly after the QCD phase transition. On the x-axis we plot the mass of the lightest doublet $m_H$ or of the lightest triplet $m_T$.}
\label{fig:LQCD2}
\end{center}
\end{figure}

These results are summarized in Fig.~\ref{fig:LQCD1} for universes with VEVs and in Fig.~\ref{fig:LQCD2} for universes without VEVs. In these Figures and in all our numerical results, when we write $\langle H \rangle$ we mean $\langle H \rangle=\langle H_u \rangle \simeq \langle H_d \rangle/\epsilon\simeq \sqrt{\langle H_u \rangle^2+\langle H_d \rangle^2}$ and similarly for $\langle T \rangle=\langle T_u \rangle \simeq \langle T_d \rangle/\epsilon$. As discussed in Section~\ref{sec:QCD} we take SUSY breaking to be approximately $SU(5)$-symmetric in the doublets and triplets sectors, so that $\epsilon$ is the same for both.

Going back to the $\phi_\pm$ potentials, we see that when the doublets have a VEV (left panel of Fig.~\ref{fig:LQCD1}) $\Lambda>\Lambda_{\rm SM}$ (unless one tunes more to get a lighter doublet than observed at the LHC) and $\phi_-$ crunches away theses unwanted universes. When the triplets have a VEV (right panel of Fig.~\ref{fig:LQCD1}) or nobody has a VEV (Fig.~\ref{fig:LQCD2}), $\Lambda \ll \Lambda_{\rm SM}$ and $\phi_+$ crunches away these universes.

We can understand qualitatively the results in Figs.~\ref{fig:LQCD1} and~\ref{fig:LQCD2} in the following way. The scale $\Lambda$ of the axion-like potentials is dominated by the IR dynamics of QCD in most universes, $\Lambda^4 \sim m_\pi^2 f_\pi^2$ for $m_{u,d}\lesssim 4\pi f_\pi$ and $\Lambda^4 \sim \LQCD^2 f_\pi^2$ otherwise, as explained in detail in Section~\ref{sec:potential}. In the left panel of Fig.~\ref{fig:LQCD1}, when the doublet VEVs are larger than the weak scale, quarks have a larger mass than in our universe and are integrating out earlier making the QCD $\beta$-function large and negative in a broader energy range, and leading to an increased confinement scale, $\Lambda > \Lambda_{\rm SM}$. In the right panel of Fig.~\ref{fig:LQCD1}, we show the results for the case with a large triplet VEV. Turning on a triplet VEV leads to the breaking of the  $SU(3)_c$ gauge group to $SU(2)_c$. The smaller group has a less negative $\beta$-function, leading to  $\LQCD$ smaller than in our universe. We see that $\Lambda$ grows with $\langle T \rangle$, because the quark masses increase with the triplet VEV and this increases $\Lambda$, just as it happens in universes with large doublet VEVs.

In Fig.~\ref{fig:LQCD2} there are neither doublet nor triplet VEVs and all SM fermions are lighter than $\LQCD$. Therefore, the QCD scale does not depend appreciably on the doublet or triplet masses. However, the scale of the axion-like potential $\Lambda$ does. We can understand this by first considering the contribution to $\Lambda$ from the strong dynamics and focus on the left panel of Fig.~\ref{fig:LQCD2}. In the left panel, $T_{u,d}$ are both at the GUT scale and do not contribute significantly to the $\phi_\pm$ potentials. As explained in Section~\ref{sec:QCD}, the pions get masses from a dimension 6 operator,  shown in Eq.~\eqref{eq:HnoVEV}, suppressed by the doublet masses $\sim B\mu/(m_U^2 m_D^2)$. So, if we call $m_H$ the mass of the lightest doublet, $\Lambda^4 \sim 1/(m_U^2 m_D^2) \sim 1/m_H^4$ when both doublets are above $\MSUSY$, and $\Lambda^4 \sim 1/(m_H^2\MSUSY^2)$ when one is tuned below, since the other cannot be lighter than $\MSUSY$, as explained in the previous Section. If $m_H \lesssim \Lambda_{\rm QCD}$ we cannot integrate out the doublets before matching the theory to the Chiral Lagrangian and there is no more $1/m_H^2$ suppression in the pion masses. In this regime we see $\Lambda$ in Fig.~\ref{fig:LQCD2} saturate at a value below $\Lambda_{\rm SM}$. 

When the lightest doublet has mass $100\;{\rm GeV} \lesssim m_H \lesssim \MSUSY$ we see from Fig.~\ref{fig:LQCD2} that UV instantons give the largest contribution to the $\phi_\pm$ potentials. In Section~\ref{sec:inst} we show that the instantons are dominated by instanton sizes of $\mathcal{O}(\MSUSY^{-1})$ and are independent of $m_H$ when $m_H < \MSUSY$. When $m_H > \MSUSY$, we have instead a rapid decoupling:
$\Lambda^4\sim\left(\MSUSY/m_T\right)^9$, as discussed in Section~\ref{sec:inst}. In the right panel of Fig.~\ref{fig:LQCD2}, $H_{u,d}$ are at the GUT scale. The physics is analogous to the left panel with the role of doublets and triplets interchanged.

In conclusion, Fig.~\ref{fig:LQCD2} shows that also in universes with no VEVs $\Lambda \ll \Lambda_{\rm SM}$.
We now turn to making the mechanism outlined in this Section more precise. In Section~\ref{sec:sliding} we describe in more detail the $\phi_\pm$ dynamics during the history of the universe and how this will select our observed universe. In Section~\ref{sec:potential} we present the computation of the $\phi_\pm$ potentials.

\section{Sliding Models}\label{sec:sliding}

In this Section we review how $\phi_\pm$ can select a small, but non-zero value for the weak scale and the QCD $\theta$-angle, following the discussion in~\cite{TitoDAgnolo:2021nhd, TitoDAgnolo:2021pjo}.
We will show that when embedded into a GUT the same dynamics selects universes with the triplet at the GUT scale and a single Higgs doublet at the weak scale that breaks the EW symmetry.

We take $\phi_\pm$ to have an axion-like coupling to $SU(5)$, $V_{H \phi}$, and a SM-independent potential $V_\pm$, as in the previous Section,
\be
V=V_\pm+V_{H \phi} =\mp\frac{m_{\phi_\pm}^2}{2}\phi_\pm^2-\frac{m_{\phi_\pm}^2}{4 M_{\pm}^2}\phi_\pm^4- \frac{\alpha_5}{8\pi}\left(\frac{\phi_+}{F_+}+ \frac{\phi_-}{F_-}+\bar \theta \right){\rm Tr}[F_5\widetilde F_5]+...
\label{eq:Vtot}
\ee
This form of the potential is valid up to $|\phi_\pm|\simeq M_\pm\ll F_\pm$, beyond this scale $V_\pm$ needs to be UV-completed as in~\cite{TitoDAgnolo:2021nhd, TitoDAgnolo:2021pjo}. This model does not pose the same problems to a UV completion as the relaxion~\cite{Gupta:2015uea, McAllister:2016vzi}, since the period of the compact part of the potential is large compared to the field excursions in the non-compact one: $F_\pm \gg M_\pm$, implying that in principle these fields can potentially be true axions.

In universes similar to our own, the dominant contribution to the potential\footnote{It can be calculated from the chiral Lagrangian.} from $V_{H \phi}$ turns on at the QCD phase transition, after chiral symmetry breaking,
\be
V_{H \phi} &\simeq& - m_\pi^2 f_\pi^2 \sqrt{1-\frac{4 m_u m_d}{(m_u+m_d)^2}\sin^2\left(\frac{\phi_+}{2 F_+}+ \frac{\phi_-}{2 F_-}+\frac{\bar\theta}{2}\right)} \nn \\
&\simeq& \frac{\Lambda^4}{2} \left(\frac{\phi_+}{F_+}+ \frac{\phi_-}{F_-}+\bar\theta\right)^2 +... \label{eq:phiH}
\ee
In the second line we have zoomed in near $|\phi_\pm|\lesssim M_{\pm}$ and taken $M_\pm/F_\pm \ll 1$. The characteristic size of the potential is
\be
\Lambda^4 = m_\pi^2 f_\pi^2 \frac{m_u m_d}{(m_u+m_d)^2} \, \quad {\rm for} \quad m_{u,d}\lesssim 4 \pi f_\pi\, . \label{eq:LQCD}
\ee
We stress that $\Lambda^4$ is a monotonic function of the Higgs VEV even in the regime $m_{u,d} \gtrsim 4 \pi f_\pi$, although the form of the potential and of $\Lambda$ becomes different. In general $\Lambda^4$ depends on the parameters scanning in the Multiverse, in our case $\Lambda^4=\Lambda^4(\mu_5)$, as in the previous Section.

Notice that Eq.~\eqref{eq:phiH} does not impose constraints on the naturalness of $V_\pm$, i.e. the coupling to $F_5 \widetilde F_5$ does not give radiative corrections to $V_\pm$ sensitive to the cutoff of the theory\footnote{More precisely they are suppressed by the instanton factor $\sim e^{-\frac{2\pi}{\alpha_5(\Lambda_{\rm UV})}}$}. As stated above, we make the technically natural choice: $M_\pm/F_\pm~\ll~1$, so that $V_{H\phi}$ is dominated by the quadratic term in the second equality of Eq.~\eqref{eq:phiH} when $|\phi_\pm|\lesssim M_{\pm}$. As we show in the following, this is required by current measurements of the QCD $\theta$-angle. It is also a technically natural choice, reminiscent of usual QCD. The scales $M_\pm$ are associated to the breaking of a shift-symmetry on $\phi_\pm$, while the couplings suppressed by $F_\pm$ preserve it. $F_\pm$ plays a role similar to $f_\pi$ in QCD, while $M_\pm$ are more similar to the quark masses.

To simplify the analysis we can assume that $\phi_-$ starts rolling before $\phi_+$ in an expanding Universe, $m_{\phi_-}\gtrsim m_{\phi_+}$. When $m_{\phi_+}\lesssim H \lesssim m_{\phi_-}$, $\phi_+$ is frozen and we can focus on the potential
\be
V \supset \frac{m_{\phi_-}^2}{2}\phi_-^2-\frac{\lambda_-}{4}\phi_-^4+\Lambda^4 \theta_{\rm eff} \frac{\phi_-}{F_-}\, , \label{eq:phi-pot}
\ee
where $\theta_{\rm eff} = \bar\theta+ \langle \phi_+ \rangle/F_+$. The typical value of $\theta_{\rm eff}$ is $\simeq \bar\theta + M_+/F_+$ because universes where the initial position of $\phi_+$ is $|\phi_+|\gg M_+$ always crunch, as we show below, and we can ignore them. For the same reason, the typical value of $\phi_-$ is $M_-$ in interesting universes, i.e. those that are not doomed to crunch from the start. 

In Eq.~\eqref{eq:phi-pot} we took $(M_-/F_-)^2$ small enough that we can ignore the $\phi_-^2$ term in the potential proportional to $\Lambda^4$.
From Eq.~\eqref{eq:phi-pot} we can conclude that a minimum for $\phi_-$ only exists if
\be
\Lambda^4 \lesssim \frac{m_{\phi_-}^2M_- F_-}{\theta_{\rm eff}} \simeq \frac{m_{\phi_-}^2M_- F_-}{\left(\bar\theta + M_+/F_+\right)}\, . \label{eq:phi-}
\ee
If this inequality is not satisfied, then the linear term in the QCD potential tilts $V_\pm$ and destroys the local minimum. In other words, in a stable universe the tadpole from the QCD potential contributes to the total potential less than the energy difference between the two maxima and the minimum. 
Since $\Lambda^4$ is monotonic in $\langle H \rangle$, this will impose an upper bound on the maximal Higgs VEV $\langle H \rangle$ in all universes where the triplet does not have a VEV. In our setup this is approximately a bound on the VEV of the lightest SM-like Higgs, as discussed in Section~\ref{sec:model}. By fixing the mass of $\phi_-$ as we show later in Eq.~\eqref{eq:mphi}, the upper bound reproduces the observed value $\langle H \rangle \simeq 174$~GeV. The other Higgs doublet is heavier and has a negligibly small VEV. Universes with a larger $\langle H \rangle$ have a larger tadpole $\Lambda^4(\langle H\rangle) \theta_{\rm eff} \phi_-/F_-$ that destroys the local minimum of $\phi_-$. To summarize, $\phi_-$ takes care of all universes where the Higgs doublets have a VEV much larger than the observed one and $\Lambda^4 \gg \Lambda^4_{\rm SM}$. 

In universes where the local minimum of $\phi_-$ is absent, it will take the  $\phi_-$ field a time of $\mathcal{O}(1/m_{\phi_-})$ to slide to a deeper minimum and crunch the universe~\cite{TitoDAgnolo:2021nhd}. If $\phi_-$ starts its life far away from its local minimum in the region $|\phi_-|\lesssim M_-$, the sliding dynamics can be slightly more complicated and depending on the initial conditions it might also lead to universes with the wrong Higgs VEV that inflate rapidly and become empty instead of crunching. This is discussed in more detail in~\cite{TitoDAgnolo:2021nhd, TitoDAgnolo:2021pjo}. However all universes where $\langle H \rangle$ is large enough to violate Eq.~\eqref{eq:phi-} do not contain observers, in the sense of Weinberg's anthropic argument for the Cosmological Constant (CC)~\cite{Weinberg:1987dv}. 

In universes where the initial position of $\phi_+$ is tuned to give $\bar\theta+\phi_{+}^{\rm initial}/F_+\simeq 0$, initially $\phi_-$ does not see a tadpole and does not roll away from its local minimum. However, after $\phi_+$ starts moving, $\phi_-$ will see an effective $\theta$-angle of $\mathcal{O}(M_+/F_+)$ and will start rolling towards the demise of the universe. This happens because $\phi_+$ moves by about $M_+$ after the QCD phase transition\footnote{There is an exception to this argument. In universes that are doubly tuned, i.e. $\theta+\phi_{+}^{\rm initial}/F_+\simeq 0$ and $\phi_{+}^{\rm initial} \simeq \phi_{+,{\rm min}}$ the $\phi_-$ tadpole can be smaller than $\Lambda^4 M_+/(F_-F_+)$, since in this case $\phi_+$ essentially does not move throughout its cosmological history. However, the two tunings needed to achieve this are more severe than the usual tunings in the SM from the Higgs hierarchy and the strong CP problem. This double tuning will make such  patches extremely rare and render them irrelevant, as discussed in detail in \cite{TitoDAgnolo:2021nhd}.}. 

At a later time ($H(T_*)\simeq m_{\phi_+}$), $\phi_+$ starts moving. At this point either the universe has already crunched (or is about to crunch if we are in the tuned case described in the previous paragraph) or $\phi_-$ is already oscillating around its local minimum, $\phi_- \sim M_-$. Therefore, at this stage we need to consider the potential
\be
V \supset -\frac{m_{\phi_+}^2}{2}\phi_+^2-\frac{\lambda_+}{4}\phi_+^4+\Lambda^4\left( \theta_{\rm eff}^\prime \frac{\phi_+}{F_+}+\frac{\phi_+^2}{2 F_+^2}\right)\, ,\label{eq:phi+pot}
\ee
where $\theta_{\rm eff}^\prime = \bar\theta+ \langle \phi_- \rangle/F_-$.
The above potential has a safe local minimum for $\phi_+$ only if
\be
\Lambda^4 \gtrsim m_{\phi_+}^2 F_+^2 \quad {\rm and} \quad \frac{M_+}{F_+} \gtrsim \theta^\prime_{\rm eff} \simeq \bar{\theta} + \frac{M_-}{F_-} \, . \label{eq:phi+}
\ee
This puts an upper bound on the maximal neutron EDM that we can observe today and crunches most universes in our Multiverse (all those without VEVs or where the triplet has a VEV). 
In these universes $\Lambda\ll \Lambda_{\rm SM}\simeq 80$~MeV, as shown in Figs.~\ref{fig:LQCD1} and~\ref{fig:LQCD2}, and $\phi_+$ is not stabilized.

 The first inequality in Eq.~\eqref{eq:phi+} means that the positive curvature from $V_{H \phi}$ beats the negative one from $V_{\phi_+}$, and is not satisfied if the Higgs VEV is too small. The second one enforces that $V_{H \phi}$ has a minimum in the flat region of $V_+$. If the minimum of $V_{H \phi}$ is shifted to a region where $V_+$ is dominated by the negative quartic $-\lambda_+ \phi_+^4$, a minimum for the full potential in Eq.~\eqref{eq:phi+pot} is never generated. In general $V_{H \phi}$ can be a sum of multiple contributions with different phases, but in universes similar to our own, where the potential is dominated by QCD, our inequality is selecting a small $\bar \theta$ entering the neutron EDM. 

We have found that $\phi_+$ gives an upper bound on $\bar\theta$ and a lower bound on $\Lambda^4$ given by Eq.~\eqref{eq:phi+}, while $\phi_-$ gives the upper bound on $\Lambda^4$ in Eq.~\eqref{eq:phi-}. The two scalars, combined, crunch all universes different from our own. In terms of experimentally measured quantities, this means that $M_-/F_-\lesssim M_+/F_+ \simeq \theta_0 \lesssim \theta_{\rm exp} \simeq 10^{-10}$, where $\theta_0$ is the observed $\theta$-angle today and $\theta_{\rm exp}$ is given by the experimental upper bound on the neutron EDM~\cite{Abel:2020pzs}. Note that we require $M_-/F_-\lesssim M_+/F_+ $ in order to be able to neglect the $\phi_-^2$ term in the QCD part of the $\phi_-$ potential in Eq.~\eqref{eq:phi-pot} and ensure that the tadpole does destroy the local minimum of $\phi_-$.
 The electroweak scale is successfully selected if
\be
m_{\phi_+}^2 &\simeq& \frac{\Lambda_{\rm SM}^4}{F_+^2}\, , \nn \\
m_{\phi_-}^2 &\simeq& \left(\theta + \frac{M_+}{F_+}\right)\frac{\Lambda_{\rm SM}^4}{ F_- M_-} \, \gtrsim \, \frac{\Lambda_{\rm SM}^4}{ F_-^2} \, , \label{eq:mphi}
\ee
where $\Lambda_{\rm SM}^4\simeq (80~{\rm MeV})^4$ is the scale in Eq.~\eqref{eq:LQCD}, and we have used $M_- \lesssim M_+ F_-/F_+$.

From bounds on axion-like particles we know that $F_\pm \gtrsim 10^8$~GeV~\cite{ParticleDataGroup:2024cfk}, if we take also $M_\pm \gg m_{\phi_\pm}$, the scalars $\phi_\pm$ are very weakly coupled (i.e. have an approximate shift symmetry), $\lambda_\pm \ll 1$, and can be naturally light. The model described here and originally presented in~\cite{TitoDAgnolo:2021nhd} is in some sense outsourcing the Higgs hierarchy problem to this weakly coupled sector where a large hierarchy of scales (between $m_{\phi_\pm}$ and $M_\pm$) is natural.  If this model is realized in Nature, measuring the weak scale in our universe is equivalent to measuring $m_{\phi_\pm}$. This is not so different from dimensional transmutation. The QCD scale depends exponentially on the strong coupling at some large UV scale, but nobody thinks of this as a tuning or a coincidence, measuring $\LQCD$ is equivalent to measuring the strong coupling in the UV.

In this Section we have not spent too much time on the crunching dynamics, since it is already described in detail in~\cite{TitoDAgnolo:2021nhd, TitoDAgnolo:2021pjo}. However it is useful to point out that the crunching time depends on initial conditions. Universes that start their life with either $|\phi_-| \gg M_-$ or $|\phi_+|\gg M_+$ crunch rapidly (or never exit inflation), regardless of the value of $\theta$ and $\langle h \rangle$. The longest-lived universes are those that start their life with $|\phi_\pm |\lesssim M_\pm$ and small kinetic energy for both scalars. In these cases the crunching dynamics starts either when $m_{\phi_+}\gtrsim H(T)$ (so right below the temperature of the QCD phase transition in our universe) or earlier, when $m_{\phi_-}\gtrsim H(T)$. Note that we need $m_{\phi_+}\lesssim H(\LQCD^{\rm SM})$ to avoid that also good universes crunch. Therefore it takes the crunching scalar a time of order $1/m_{\phi_\pm} \lesssim 1/H(\LQCD^{\rm SM}) \simeq 10^{-5}$~s to reach the deep minimum. As explained in more detail in~\cite{TitoDAgnolo:2021nhd}, the time needed to crunch is dominated by the dynamics near the origin $|\phi_{\pm}|\lesssim M_{\pm}$ and it takes only an additional negligibly short time to reach a deeper minimum from $|\phi_{\pm}|\simeq M_{\pm}$.

Incidentally we note that the flat direction of the QCD term in Eq.~\eqref{eq:phiH} does not pose a problem for the stability of the potential, independently of the Higgs VEV. It is easy to show that for $M_-/F_- \lesssim \theta_0$ the flat direction of Eq.~\eqref{eq:phiH} is stabilized by $V_-$~\cite{TitoDAgnolo:2021nhd}.

\subsection*{Cutoff and Field Excursions}

In the previous discussion we did not fix the value of the Higgs cutoff scale; indeed it could take any value and does not have to be related to $M_\pm$ or $F_\pm$. Let us consider for concreteness a model that is easy to embed in a GUT. Take $F_\pm \gtrsim \MGUT$ and SUSY broken at an intermediate scale, $\widetilde M_S\simeq 10^6$~GeV, that, as we discussed above, is a good benchmark for unification. 
The Higgs and CC naturalness cutoffs are at the SUSY breaking scale $\widetilde M_S$, below $F_\pm$ and $\MGUT$. The scale $M_\pm$ can be anywhere, in principle it could also be well below $\widetilde M_S$. To highlight the parametrics we take the scales of the $\phi_\pm$ potentials to be equal in the remainder of this Section and call $M_*$ the common value of $M_\pm$.

The maximal value of the CC in this model, $\Lambda_{\rm max}\sim \Lambda_S$, fixes the position of the deep minima of the $\phi_\pm$ potentials. If we replace Eq.~\eqref{eq:pot1} with the UV-complete potential
\be
V_\pm=m_{\phi_\pm}^2 M^2_* \left(\frac{\phi_\pm}{M_*} +\frac{\phi^2_\pm}{2 M^2_*}\pm\frac{\phi^3_\pm}{3 M^3_*} + \frac{\delta}{4} \frac{\phi^4_\pm}{M^4_*}\right)+... \, , \label{eq:V1}
\ee
we have a deep minimum at a distance 
\be
\Delta \phi \simeq \frac{M_*^{1/3}\Lambda_S^{4/3}}{m_\phi^{2/3}} \simeq M_*^{1/3} F^{2/3}\left(\frac{\Lambda_S}{\LQCD}\right)^{4/3} \gtrsim 10^{20} M_*\, 
\ee
from the origin. In the last inequality we have used $M_*/F \lesssim 10^{-10}$, needed to select the observed $\theta$-angle. In the example potential of Eq.~\eqref{eq:pot1} instead we have
\be
\Delta \phi \simeq \Lambda_S \sqrt{\frac{M_*}{m_\phi}}\simeq \frac{\Lambda_S}{\LQCD} \sqrt{M_* F} \gtrsim 10^{15} M_*\, . 
\ee
The difference is due to the different powers of $\phi_\pm$ in the two potentials. In Eq.~\eqref{eq:V1} the $\phi^3_\pm$ terms dominate the approach to the deep minimum, while in the above equation it is the $\phi^4_\pm$ terms that dominate. 

If we want the model to satisfy the swampland distance conjecture~\cite{Ooguri:2006in} that implies $\Delta \phi \lesssim c M_{\rm Pl}$, with $c$ an $\mathcal{O}(1)$ number, we have to take $M_*$ rather small, but this does not pose any problem to the mechanism. The simplest way to select the weak scale is to imagine that the sector generating $V_\pm$ is sufficiently cold at reheating, so when the SM photons are at $T\simeq \LQCD$, $V_\pm$ already exists in its zero-temperature form. For a SM reheating temperature sufficiently below $F_\pm \gtrsim \MGUT$ the $\phi_\pm$ sector is not appreciably reheated by the SM.

We can make this assumption on the reheating temperature of $\phi_\pm$ to show the existence of a concrete scenario where the Sliding Naturalness mechanism works and respects the distance conjecture. This choice also simplifies the discussion in the previous Section, but it is not necessary for the mechanism to work. 

\section{Axion-Like Potentials}\label{sec:potential}

We have explained above how the Sliding Naturalness mechanism can be implemented in GUTs, and how it will also solve the doublet-triplet splitting problem. There is one important remaining task which we undertake in this Section: so far we showed that with the potential \eqref{eq:Vtot} the Sliding Naturalness mechanism will choose universes with particular values of $\Lambda^4$ and $\bar\theta$, which can agree with those observed in the SM by appropriate choices of the parameters. However one still has to ensure that in GUTs there are no  dynamical effects that would generate additional terms for the $V_{H\phi}$ potential, which could keep unwanted universes from crunching.

The $\phi_\pm$ potentials are dominated by low energy contributions from the QCD strong dynamics in most of our universes. UV instanton contributions are small compared to $\Lambda_{\rm SM}^4$. As we will discuss in Section~\ref{sec:inst}, they are at most of order 
\be
V_{\rm inst} \sim \frac{12\;\epsilon^3 \MSUSY^3 \MGUT}{(8\pi^2)^4} \left(\frac{2\pi}{\alpha_s(\MSUSY)}\right)^6 \det(Y_u Y_d)   e^{-\frac{2\pi}{\alpha_s(\MSUSY)}}\, . 
\ee
{They are suppressed by the small gauge coupling at high energies $e^{-2\pi/\alpha}$, by powers of the SUSY-breaking scale $\MSUSY$, and by $\det(Y_u Y_d)$. More details, and the selection rules that lead to these suppression factors, are given in Section~\ref{sec:inst}. Numerically, }
for our benchmark point\footnote{We derive our benchmark point by running the SM and MSSM RGEs with all superpartners at $\MSUSY$ and finding good unification with the values of $\MGUT$ and $\aGUT$ in the main text.}: $\MSUSY=10^6$~GeV, $\MGUT\simeq1.3\times10^{16}$~GeV, 
$\aGUT\simeq(25.3)^{-1}$, {we find $V_{\rm inst}^{1/4}/\Lambda_{\rm SM}\lesssim 10^{-4}.$} Lowering the scale of SUSY breaking (for instance taking $\MSUSY= 5$~TeV, $\MGUT\simeq 10^{16}$~GeV, $\aGUT\simeq (26.2)^{-1}$) makes $V_{\rm inst}$ even smaller. Instantons are always negligible compared to $\Lambda^4_{\rm SM} \simeq (80\;{\rm MeV})^4$, {which makes them negligible compared to $V_\pm$, the part of the $\phi_\pm$ potential that does not originate from the $\phi_\pm$ interactions with the SM. As discussed in the previous Section we take $V_\pm$ to be fixed throughout the Multiverse. Therefore, instantons cannot stabilize $\phi_+$} and cannot prevent a universe from crunching.

In the next two Sections we compute in detail both classes of contributions to the total $\phi_\pm$ potentials (IR strong dynamics and UV instantons) to show that they can be comparable to our universe only in universes that are qualitatively similar to it. 

\subsection{IR Strong Dynamics}\label{sec:QCD}
QCD confines in all our universes, either as $SU(3)_c$ or $SU(2)_c$. We find that the QCD scale is always larger than the scale at which $SU(2)_L$ would have confined in universes where $\langle H \rangle =0$. Therefore, when computing the low energy contributions to the $\phi_\pm$ potentials, we can focus on QCD alone, and ignore $SU(2)_L$ that is broken by either $\langle H\rangle$ or the QCD condensate before confining.

\paragraph{$\mathbf{\left<H \right> >0 }$ and $\mathbf{\left<T \right> =0 }$} In universes where the doublets have a VEV, but the triplets do not, the calculation is the same as in our universe. From the Chiral Lagrangian we obtain
\be
\Lambda^4(\langle H \rangle, 0) =m_\pi^2 f_\pi^2 \frac{m_u m_d}{(m_u+m_d)^2} \quad \text{if \quad $4\pi f_\pi \gtrsim m_{u,d}$}\, . \label{eq:L1}
\ee 
To improve readability, in this Section we show explicitly the dependence of $\Lambda^4$ on the doublets and triplets VEVs, $\Lambda^4=\Lambda^4(\langle H \rangle, \langle T\rangle)$.
To estimate Eq.~\eqref{eq:L1} we compute the strong coupling scale at one loop
\be
\LQCD = \mu e^{-\frac{8 \pi^2}{b_0(\mu) g^2_3(\mu)}}\, ,
\ee
and take $f_\pi = \LQCD/4\pi$, $m_\pi^2 = c (m_u+m_d)\LQCD$, with $c\simeq 0.6$ that reproduces the neutral pion mass in our universe. We compute the one-loop $\beta$-function, $b_0$, in the $\overline{\text{MS}}$ scheme and decouple heavy particles at their mass threshold. 

In universes without light quarks ($y_u \langle H_u \rangle > 4\pi f_\pi$), $\phi_\pm$ have a potential that we can estimate using NDA
\be
\Lambda^4(\langle H \rangle, 0) \simeq \alpha \LQCD^2 f_\pi^2 \quad \text{if \quad $4\pi f_\pi < m_{u,d}$}\, . \label{eq:L2}
\ee
We fix $\alpha$ by requiring a continuous transition to the result from the Chiral Lagrangian in Eq.~\eqref{eq:L1}.

\paragraph{$\mathbf{\left<H \right> =0 }$ and $\mathbf{\left<T \right> >0 }$} In universes where the triplets have a VEV, strong interactions below $\langle T \rangle$ are described by a $SU(2)_c$ gauge symmetry. All quarks charged under $SU(2)_c$ get a mass from the triplet VEV, as discussed below. 

The flavor symmetry of a $SU(2)_c$ gauge theory with $F$ flavors is $SU(2F)$. The reason is simple: the fundamental representation of $SU(2)$ is pseudo-real and there is no distinction between $q$ and $q^c$. The QCD condensate breaks the flavor symmetry to $USp(2F)$~\cite{Peskin:1980gc}. Hence one has a slightly different symmetry breaking pattern here than in ordinary QCD, implying that the form of the Chiral Lagrangian will need to be adjusted to this new symmetry breaking pattern $SU(2F)/USp(2F)$. The leading terms in this new Chiral Lagrangian are
\be
\mathcal{L}_{\pi\, SU(2)} = \frac{f^2_\pi}{4}{\rm Tr}[(\partial_\mu \Sigma)^\dagger \partial_\mu \Sigma] -\frac{f^2_\pi \LQCD}{4}\textnormal{Tr}\left[\Sigma\Omega\mathcal{M}+\textnormal{h.c.} \right]\, , \label{eq:cL2}
\ee
where $\mathcal{M}$ is the quark mass matrix and
\be
\Sigma(x)&=&\exp\left(i\frac{\pi(x)}{f_\pi} \right)\Omega\exp\left(i\frac{\pi(x)^T}{f_\pi} \right)\, , \quad \pi(x) \equiv \sum_a \pi^a(x) X^a\, , \nn \\
\Omega&=&\begin{pmatrix}
0 && \mathds{1}_{F\times F}\\
-\mathds{1}_{F\times F} && 0\\
\end{pmatrix}\, .
\ee
The generators $X^a$ are those in the coset $SU(2F)/USp(2F)$. In the following we give more details on the Yukawa couplings entering the quark mass matrix $\mathcal{M}$ in Eq.~\eqref{eq:cL2}, but first we note that we can compute the axion mass as in~\cite{Csaki:2023yas}, following the same procedure as for the ordinary Chiral Lagrangian in the SM, and  obtain a very similar result
\be
\Lambda^4(0, \langle T \rangle) = \frac{1}{2}\Lambda^4(\langle H \rangle, 0)\, ,
\ee
with $\Lambda^4(\langle H \rangle, 0)$ in Eq.~\eqref{eq:L1}, when the up and down quarks are lighter than the confinement scale. If all quarks are heavier than the confinement scale, we use $\Lambda^4(\langle H \rangle, 0)$ given by Eq.~\eqref{eq:L2}. The main difference compared to the SM is that quark masses in these universes come from the Lagrangian  
\be
\mathcal{L} \supset -\frac{1}{2}Y_{uT} Q Q T_u + Y_{dT} u^c d^c T_d + {\rm h.c.}\, ,  \label{eq:TYuk}
\ee
where $Y_{uT},Y_{dT}$ are $3\times 3$ matrices in flavor space.
At the GUT scale $Y_{uT,dT}$ are identical to the usual doublets' Yukawa couplings $Y_u Q H_u u^c$ and $Y_d Q H_d d^c$, since they all come from the same $SU(5)$-symmetric Yukawas: $Y_{uT}=Y_u=Y_{10}$ and $Y_{dT}=Y_d=Y_{5}$. However, the running from the GUT scale down to $\LQCD$ makes them slightly different than $Y_{u,d}$ in our universe. 

As an example, we can consider the mass spectrum of the theory for $\langle T_u \rangle^2+\langle T_d\rangle^2 \simeq (100$~GeV$)^2$. The mass terms for the two lightest quarks are
\be
\mathcal{L} \supset -\frac{y_{uT} \langle T_u \rangle}{2} Q_u Q_u + y_{dT} \langle T_d \rangle u^c d^c+ {\rm h.c.}\, .  \label{eq:TYuk}
\ee
They are consistent with all the symmetries of the theory:  hypercharge is broken by the triplet VEV, but color $SU(2)_c$ and $SU(2)_L$ are unbroken. The mass term $Q_u Q_u$, with both doublets of the same flavor, is not zero, because the color $SU(2)_c$ contraction is antisymmetric.
The resulting quark masses are close to those in our universe, $m_{uT} \simeq m_u$ and $m_{dT} \simeq m_d$, if we assume 
\be
\frac{\langle T_u \rangle}{\langle T_d \rangle} \simeq \frac{\langle H_u \rangle_{\rm us}}{\langle H_d \rangle_{\rm us}}\, ,
\ee
where the subscript ``us" denotes our universe.  This amounts to assuming that SUSY breaking does not scan across the Multiverse and that it is approximately $SU(5)$-symmetric in the doublet and triplet sectors. The first assumption was already introduced in Section~\ref{sec:model} and is necessary for the mechanism to work. The assumption that $\langle T_u \rangle/\langle T_d \rangle$ is equal to $\tan\beta$ in our universe is not crucial, but it makes our usual physical intuition about the quarks' mass spectrum valid also in universes with a triplet VEV, simplifying our calculations.

When plotting Fig.~\ref{fig:LQCD1} and showing numerical results, we make these assumptions, so the mass spectrum in our universe for a given value of the Higgs VEV is approximately realized in these universes for the same value of the triplet VEV.

\paragraph{$\mathbf{\left<H \right> =0 }$ and $\mathbf{\left<T \right> =0 }$}  In these universes all SM quarks live below the QCD scale and we have to consider a Chiral Lagrangian with the non-anomalous flavor symmetry $SU(6)_L \times SU(6)_R \times U(1)_V$. The contribution of six light quarks to the QCD $\beta$-function makes the QCD scale quite small, $\Lambda_{\rm QCD}\lesssim 10^{-5}$~GeV. So, for simplicity, we assume that there are not enough vacua in the Multiverse to scan the doublet or triplet masses below such a low $\Lambda_{\rm QCD}$ and in this Section always take $H_{u,d}$ and $T_{u,d}$ parametrically heavier than the confinement scale and integrate them out before matching to the Chiral Lagrangian. In Fig.~\ref{fig:LQCD2} we show an estimate of the $\phi_\pm$ potentials when the masses are smaller. 

When the doublets are lighter than the triplets (left panel of Fig.~\ref{fig:uni2}), as a first approximation we can ignore the Yukawa couplings of the triplets, since they are integrated out at the GUT scale. The QCD condensate breaks  $SU(6)_L \times SU(6)_R$ spontaneously to its diagonal subgroup and the doublet Yukawa couplings explicitly break most of the $SU(6)_L \times SU(6)_R$ symmetry. However the $SU(2)_L$ subgroup corresponding to weak interactions is unbroken in the Lagrangian by construction, i.e. there is no explicit breaking of the gauge symmetry. When the QCD condensate breaks spontaneously the flavor symmetry subgroup $SU(2)_L \times SU(2)_R$ to the diagonal we are left with three massless pions. Since $SU(2)_L$ is gauged these three pions are eaten by $W^\pm$ and $Z$ bosons. After integrating out $H_{u,d}$, that have the same potential as in the MSSM, we find the leading operators that break $SU(6)_L \times SU(6)_R$ explicitly to $SU(2)_L$,
\be
\mathcal{L}_H \supset \left(B\mu \frac{(Y_uQ u^c)(Y_d Q d^c)}{m_U^2 m_D^2}+{\rm h.c.}\right)+\frac{|Y_u Q u^c|^2}{m_U^2}+\frac{|Y_d Q d^c|^2}{m_D^2}\, .
\label{eq:HnoVEV}
\ee
They all preserve the EW symmetry, but they break the flavor symmetries of the quark sector, including the anomalous $U(1)_A$. The operators in the above Equation can be matched to terms in the Chiral Lagrangian that give a mass to pions and $\phi_\pm$. The matching is discussed in Appendix~\ref{app:goldstones}, where we obtain the low energy effective Lagrangian
\be
\mathcal{L}_{\pi H} \supset \Lambda_{\rm QCD}^2 f_\pi^4\left(\frac{B\mu}{m_U^2 m_D^2}{\rm Tr}[Y_u^6 U Y_d^6 U]+{\rm h.c.}+\frac{|{\rm Tr}[Y_u^6 U]|^2}{m_U^2}+\frac{|{\rm Tr}[Y_d^6 U]|^2}{m_D^2}\right)\, .
\ee
Here $Y_{u,d}^6$ are $6\times 6$ embeddings of the usual Yukawa matrices, written in Eq.~\eqref{eq:Y6}.
As usual $U=\exp(i T^a \pi^a(x)/f_\pi)$, with $T^a$ the generators of $SU(6)$. Every operator comes with an unknown $\mathcal{O}(1)$ coefficient that here and in the following we set to one. 

We can compute $\Lambda^4$ as in the SM + one axion (as discussed in Appendix~\ref{app:goldstones}) and obtain, 
\be
\Lambda^4(0,0) = \Lambda_{\rm QCD}^2 f_\pi^4 \frac{B\mu}{m_U^2 m_D^2} \left(\sum_{i=1}^3 \frac{1}{y_{u_i} y_{d_i}}\right)^{-1}\left[ 1+ \mathcal{O}\left(\frac{m_\pi^2}{\Lambda_{\rm QCD}^2}\right)\right] \, .\label{eq:lightHnoVEV}
\ee
It is easy to check that, as expected, $\Lambda^4(0,0)\to 0$ if any of the Yukawa couplings goes to zero. It is straightforward to compute an equivalent expression when $m_D^2 \ll m_U^2$.

Now we can consider universes where the doublets are at the GUT scale, while the triplets are lighter, but still heavy compared to the QCD scale. The leading operators generated by integrating out the triplets are
\be
\mathcal{L}_T \supset \left(B\mu_T \frac{(Y_{uT} Q Q)(Y_{dT} u^c d^c)}{2m_{U_T}^2 m_{D_T}^2}+{\rm h.c.}\right)+\frac{|Y_{uT} Q Q|^2}{4m_{U_T}^2}+\frac{|Y_{dT} u^c d^c|^2}{m_{D_T}^2}\, , \label{eq:triplet_pions}
\ee
and they match to the Chiral Lagrangian
\be
\mathcal{L}_{\pi T} \supset \Lambda_{\rm QCD}^2 f_\pi^4\left(\frac{B\mu_T}{m_{U_T}^2 m_{D_T}^2}{\rm Tr}[Y_{uT}^6 U Y_{dT}^6 U^T]+{\rm h.c.}\right)\, ,
\ee
as discussed in Appendix~\ref{app:goldstones}. The last two operators contain four left-handed quarks or four right-handed quarks and do not match to pion operators relevant to our calculation. An axion mass calculation equivalent to the SM case and detailed in Appendix~\ref{app:goldstones} gives formally the same result as for light Higgs doublets
\be
\Lambda^4(0,0)=\Lambda_{\rm QCD}^2 f_\pi^4 \frac{B\mu_T}{m_{U_T}^2 m_{D_T}^2} \left(\sum_{i=1}^3 \frac{1}{y_{uT_i} y_{dT_i}}\right)^{-1}\left[ 1+ \mathcal{O}\left(\frac{m_\pi^2}{\Lambda_{\rm QCD}^2}\right)\right] \, .
\label{eq:lightTnoVEV}
\ee
The two scales in Eqs.~\eqref{eq:lightHnoVEV} and~\eqref{eq:lightTnoVEV} are shown in Fig.~\ref{fig:LQCD2} and are much smaller than $\Lambda_{\rm SM}$ for any choice of doublets' and triplets' masses. 

\begin{figure}[!t]
\begin{center}
\includegraphics[width=0.8\textwidth]{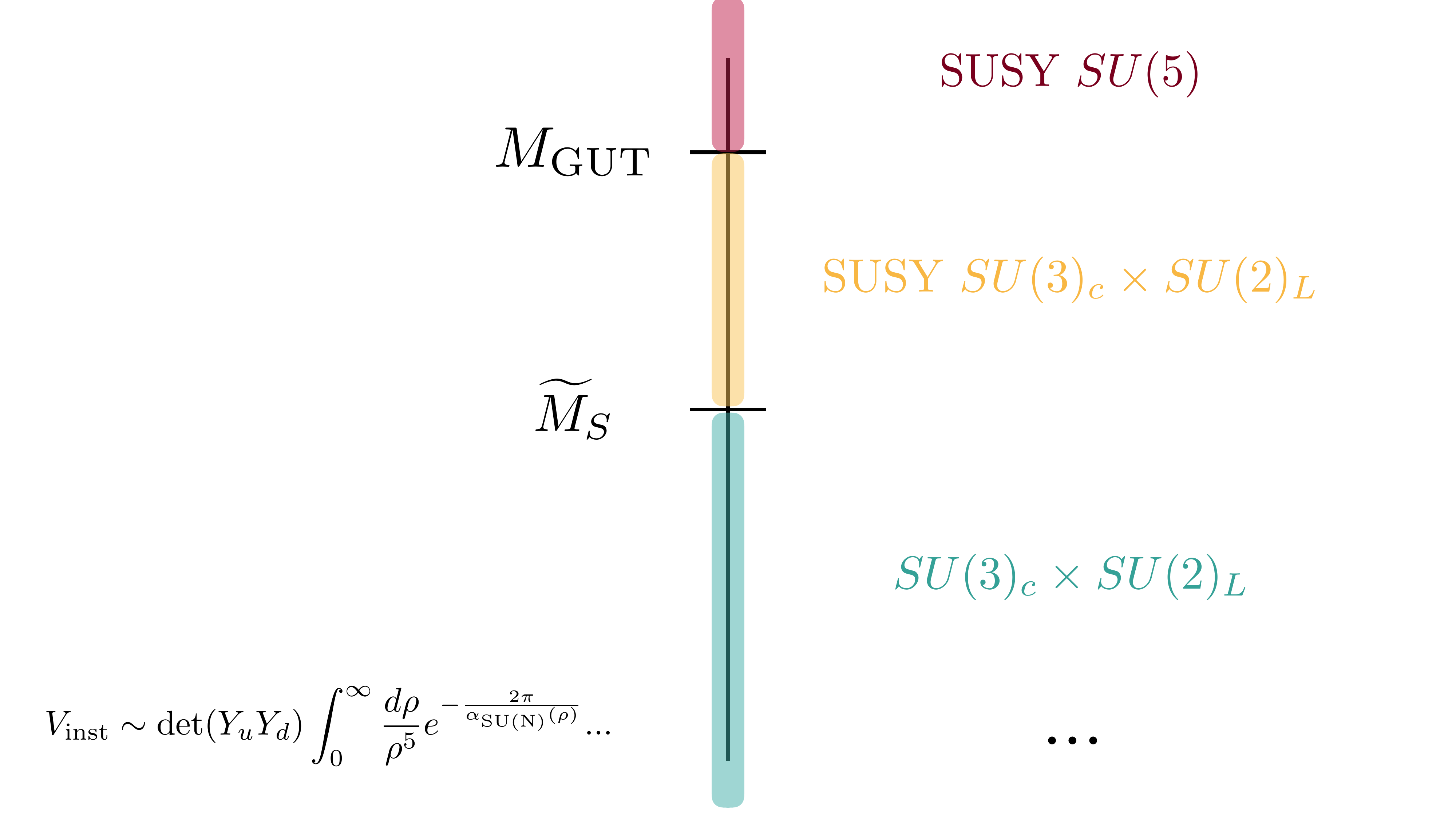}
\caption{Schematic representation of instanton contributions to the $\phi_\pm$ potentials. The potentials receive contributions from all instanton sizes $\rho$. We compute them by breaking down our model into a series of effective theories depicted in the Figure. In all universes, $SU(2)_L$ is broken at low energy, either by the doublets VEVs or by the QCD condensate. In some universes $SU(3)_c$ is broken to $SU(2)_c$ by the triplets VEVs. These two breakings add extra steps to our ladder of EFTs that we do not show in the Figure.}
\label{fig:InstSketch}
\end{center}
\end{figure}

\subsection{UV Instantons}\label{sec:inst}

\begin{figure}[!t]
\begin{center}
\includegraphics[width=0.4\textwidth]{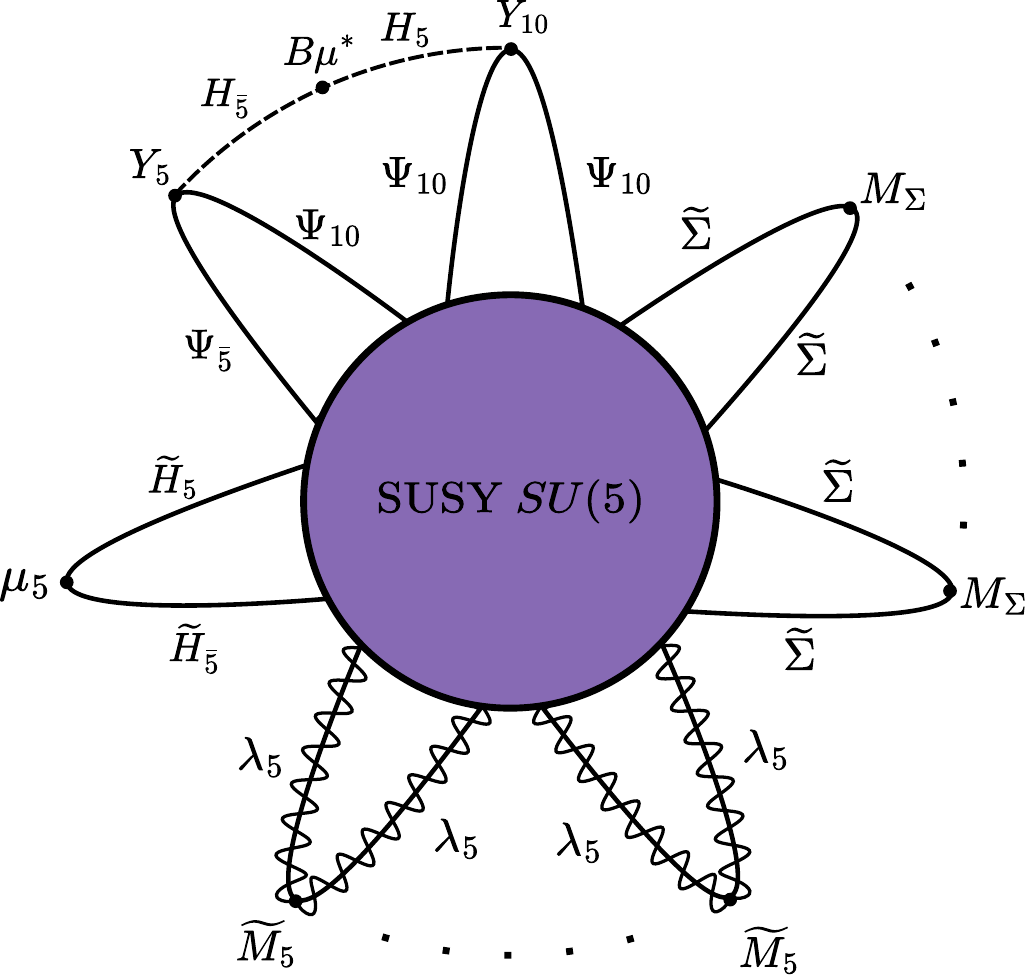}
\caption{Supersymmetric $SU(5)$ instanton contribution to the axion-like $\phi_\pm$ potentials. $\widetilde M_5$ is the gaugino mass of $\mathcal{O}(\MSUSY=10^6~\text{GeV})$. There are 10 gaugino legs and 10 $\widetilde \Sigma$ legs in the diagram from their $SU(5)$ Dynkin index that in this figure we close using fermion masses.}
\label{fig:SU5inst}
\end{center}
\end{figure}

The $\phi_\pm$ potentials receive contributions from instantons of all sizes, as depicted schematically in Fig.~\ref{fig:InstSketch}. The simplest way to compute them is to consider a series of effective theories, valid at different scales. Going from high to low energies we will discuss in order: SUSY $SU(5)$ instantons, SUSY $SU(3)_c$ and $SU(2)_L$ instantons and SM $SU(3)_c$ and $SU(2)_L$ instantons. In this Section we show the diagrams and analytical expressions for the largest contributions in each of the EFTs and only briefly comment on subleading diagrams, a more detailed discussion can be found in~\cite{Sesma:2024tcd}.

To simplify our results we introduce the field $\Theta(\phi_\pm)$, the reduced instanton density of $SU(N)$, $d_N(\rho)$, and the higgsino and tripletino masses $\mu_H$, $\mu_T$ 
\be
\Theta(\phi_\pm)&\equiv&\frac{\phi_-}{F_-}+\frac{\phi_+}{F_+}+\overline \theta\, , \\
d_N(\rho)&\equiv&\frac{2^{-2N+2}}{\pi^2(N-1)!(N-2)!}\left(\frac{8\pi^2}{g_N^2}\right)^{2N} e^{-\frac{8\pi^2}{g^2_N(1/\rho)}}\, , \\
\mu_H &\equiv& \mu_5 + 3 \lambda v_\Sigma\, , \quad \mu_T \equiv \mu_5 - 2 \lambda v_\Sigma\, . \label{eq:mudef}
\ee
In this Section we do not distinguish between a Lagrangian parameter and its complex conjugate, as we want to focus on the numerical contributions of instantons to the axion-like potentials. In Section~\ref{sec:SR} we are going to be more precise and summarize all distinct classes of instanton contributions and the selection rules that identify them.
\paragraph{SUSY Instantons} In the UV, above the GUT scale, we have a supersymmetric $SU(5)$ theory. Its leading instanton contribution to the $\phi_\pm$ potentials, depicted in Fig.~\ref{fig:SU5inst}, is the same in all universes. At leading order in $B\mu$ we have
\be
\frac{V_{\rm SU(5)}}{2\cos \Theta(\phi_\pm)}&=& 3^{n_g}(n_g!)\text{det}(Y_5Y_{10})_{2 n_g} \int_0^{\MGUT^{-1}}\frac{d\rho}{\rho^5}d_5(\rho)\rho^{2n_g}(\rho \mu_5) (\rho \MSUSY)^5 (\rho M_\Sigma)^5 \nn \\
&\times &\left[\frac{4\rho^4 B\mu }{\pi^4}\int_{\{x_i\}}\frac{D_{H_{\bar{5}}}(x_1-x_3) D_{H_5}(x_3-x_2)}{(x_1^2+\rho^2)^3(x_2^2+\rho^2)^3}\right]^{n_g}\, , \label{eq:SU5inst}
\ee
where we gave the result for an arbitrary number of generations $n_g$, and in our case $n_g=3$, so we wrote $\text{det}(Y_5Y_{10})_{2 n_g}$ to indicate that the determinant at the GUT scale contains all six flavors. The Higgs propagators in the previous expression are
\be
D_{H_{\bar{5}}}(x-y)=D_{H_5}(x-y)=\int \frac{d^4 p}{(2\pi)^4}\frac{e^{-i p\cdot (x-y)}}{p^2+i\epsilon}\, ,
\ee
because Eq.~\eqref{eq:SU5inst} is understood to be valid at energies well above their masses. Technically, this means that the integral is evaluated somewhat above the GUT scale in our universe, where bounds from proton decay place the Triplet around $10^{17}$~GeV~\cite{PhysRevD.105.112012, PhysRevD.102.112011, PhysRevD.95.012004, PhysRevD.96.012003, Murayama:2001ur}. In Appendix~\ref{app:integrals} we show how to solve the integrals in Eq.~\eqref{eq:SU5inst} analytically; these also appear in instanton contributions at lower energies.

\begin{figure}[!t]
\begin{center}
\includegraphics[width=0.4\textwidth]{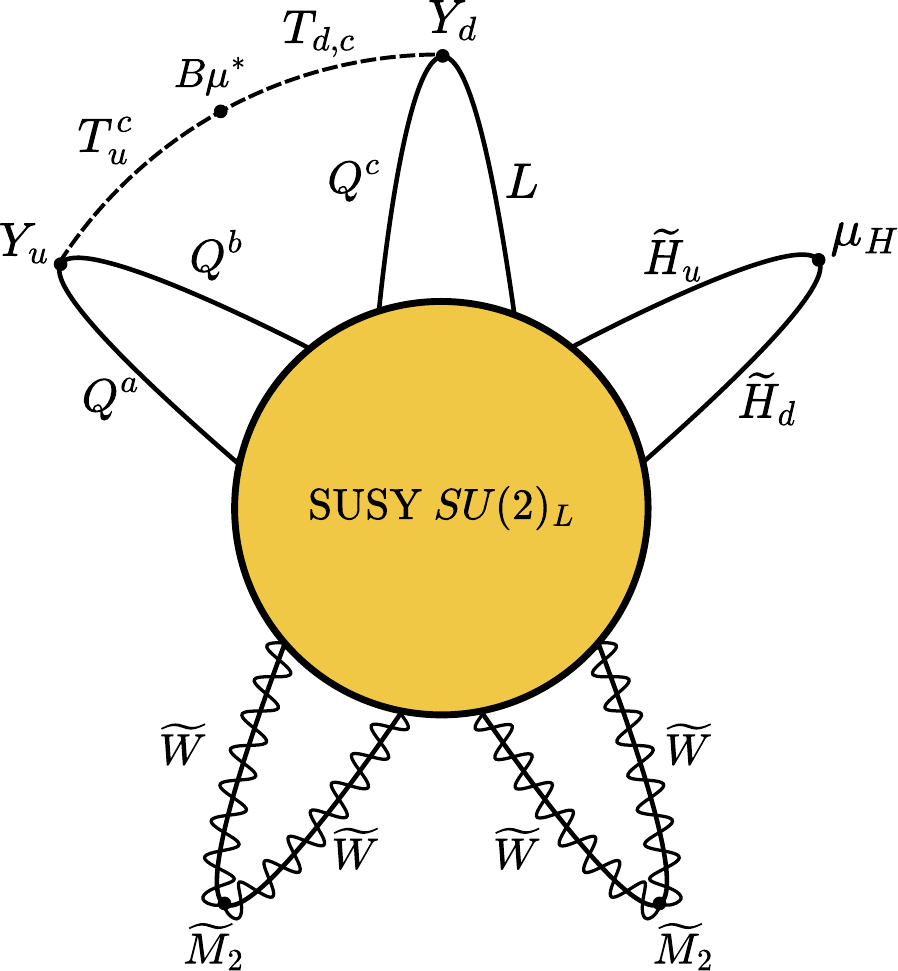}
\caption{Supersymmetric $SU(2)_L$ instanton contribution to the axion-like $\phi_\pm$ potentials. $\mu_H$ is the higgsino mass in Eq.~\eqref{eq:mudef} and $\widetilde M_2$ the wino mass. The indices on the quark doublets give an example of a possible color assignment.}
\label{fig:SUSYSU2inst}
\end{center}
\end{figure}

Eq.~\eqref{eq:SU5inst} gives, parametrically,
\be
\frac{V_{\rm SU(5)}}{\cos \Theta(\phi_\pm)}\simeq \frac{9 \; e^{-\frac{2\pi}{\aGUT}}}{2^7(8\pi^2)^4} \frac{\epsilon^3\MSUSY^{11}}{\MGUT^7}\left(\frac{2\pi}{\aGUT}\right)^{10}\text{det}(Y_5Y_{10})_6 \simeq (10^{-32}\;{\rm GeV})^4\, ,\label{eq:SU5n}
\ee
for $\MSUSY=10^6$~GeV, $\MGUT=1.3\times10^{16}$~GeV, $\aGUT=(25.3)^{-1}$ and $\epsilon=0.1$.
The main features of this expression are present in all other instanton contributions that we discuss in the following, and it is worth to comment on some of them. There is a factor of $e^{-8\pi^2/g^2}$ from the saddle point approximation of the path integral around the instanton solution of the equations of motion. We always have a factor of $\det(Y_{10} Y_5)\sim \det(Y_u Y_d)$ coming from the selection rules of the anomalous $U(1)_A$ of the SM. Additionally, there are at least two powers of a SUSY breaking scale, as discussed in~\cite{Choi:1998ep, Dine:1986bg}. 

To conclude the discussion of $SU(5)$ instantons, we note that, if both Higgses have negative squared masses around $\MSUSY$, there is a contribution larger than Eq.~\eqref{eq:SU5n} by a factor of $\epsilon^3$, but still completely negligible compared to the QCD scale in our universe.

\begin{figure}[!t]
\begin{center}
     \begin{subfigure}[b]{0.35\textwidth}
         \includegraphics[width=\textwidth]{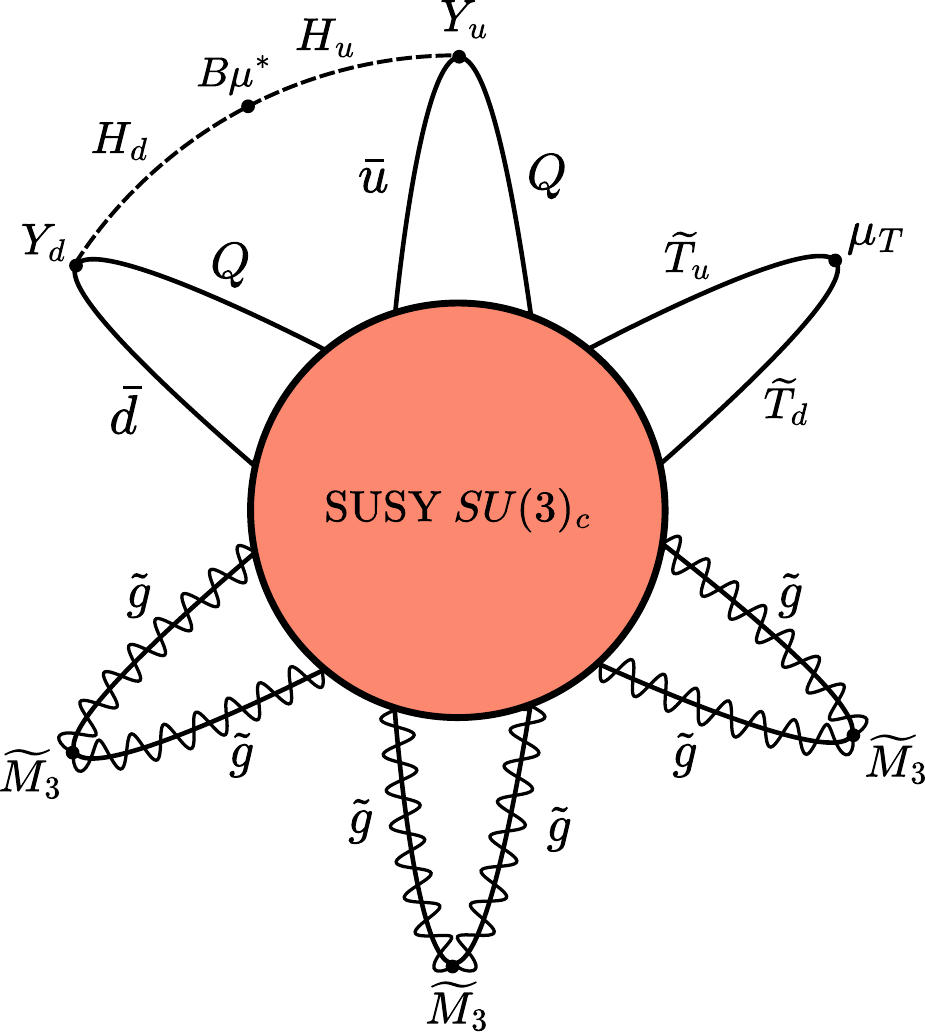}
         \caption{Doublet loops}
         \label{SU(2)Ltriplet}
     \end{subfigure}
\quad \quad \quad \quad
     \begin{subfigure}[b]{0.35\textwidth}
         \includegraphics[width=\textwidth]{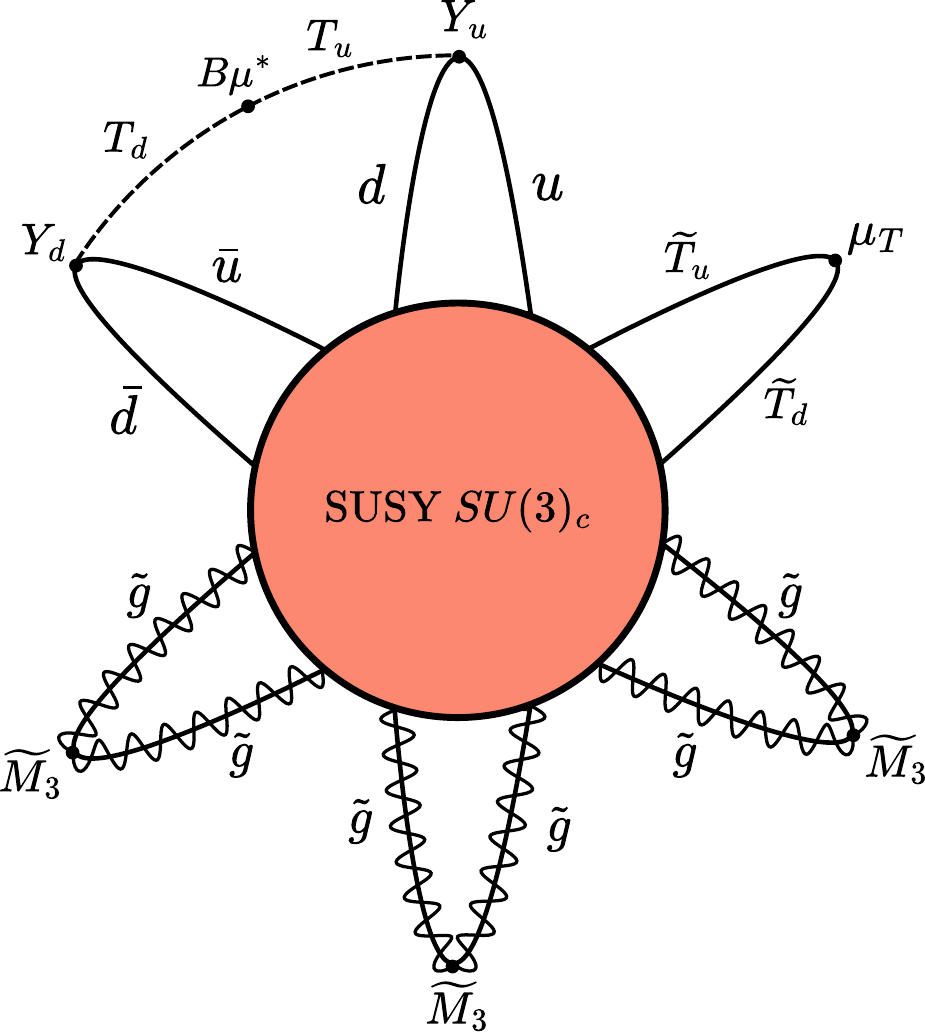}
         \caption{Triplet loops}
         \label{SU(2)Ltriplet}
     \end{subfigure}
        \caption{Supersymmetric QCD instanton contributions to the axion-like $\phi_\pm$ potentials. $\mu_H$ and $\mu_T$ are the higgsino and tripletino masses, respectively, defined in Eq.~\eqref{eq:mudef} and $\widetilde M_3$ is the gluino mass. The indices on the quark doublets give an example of a possible color assignment.}
        \label{fig:SUSYSU3inst}
        \end{center}
\end{figure}

Below $\MGUT$, but above $\MSUSY$ we have supersymmetric instantons in $SU(3)_c$ and $SU(2)_L$, depicted in Fig.s~\ref{fig:SUSYSU2inst} and \ref{fig:SUSYSU3inst},
\be
\frac{V_{\rm SU(3)_c}^{\rm SUSY}}{2\cos \Theta(\phi_\pm)}&=& 2^{n_g}(n_g!)\det(Y_u Y_d)_{2n_g}\int_{\MGUT^{-1}}^{\widetilde M_S^{-1}}\frac{d\rho}{\rho^5}d_3(\rho)\rho^{2n_g}(\rho\mu_T)(\rho\widetilde{M}_3)^3\nonumber \\
&\times&\left[\frac{4\rho^4 B\mu}{\pi^4}\int_{\{x_i\}}\frac{D_{H_u}(x_1-x_3) D_{H_d}(x_3-x_2)}{(x_1^2+\rho^2)^3(x_2^2+\rho^2)^3} \right]^{n_g} \, , \label{eq:SU3SUSY}\\ 
\frac{V_{\rm SU(2)_L}^{\rm SUSY}}{2\cos \Theta(\phi_\pm)}&=& 3^{n_g}(n_g!)\det(Y_{uT} Y_{dT})_{2n_g} \int_{\MGUT^{-1}}^{\widetilde M_S^{-1}}\frac{d\rho}{\rho^5}d_2(\rho)\rho^{2n_g}(\rho\mu_H)(\rho\widetilde{M}_2)^2 \nn \\
 &\times &\left[\frac{4\rho^4 B\mu }{\pi^4}\int_{\{x_i\}}\frac{D_{T_u}(x_1-x_3)D_{T_d}(x_3-x_2)}{(x_1^2+\rho^2)^3(x_2^2+\rho^2)^3}\right]^{n_g}\, . \label{eq:SU2SUSY}
\ee
In the $SU(3)_c$ case there is a second contribution in addition to Eq.~\eqref{eq:SU3SUSY}, which corresponds to the right panel of Fig.~\ref{fig:SUSYSU3inst}. It can be obtained by dividing Eq.~\eqref{eq:SU3SUSY} by $2^{n_g}$ and replacing doublet propagators with triplet propagators and $\det(Y_u Y_d)$ with $\det(Y_{uT} Y_{dT})$. Additionally, we can mix the two options, and close some generations with Triplet loops and others with Higgs loops. 

The masses of doublets and triplets can be important in these diagrams and we include them in the propagators $D_{H_{u,d}, T_{u,d}}$. The SUSY breaking parameters in Eqs.~\eqref{eq:SU3SUSY} and~\eqref{eq:SU2SUSY} make the integrals IR dominated and parametrically we have contributions to the $\phi_\pm$ potentials saturated at the SUSY breaking scale
\be
\frac{V_{\rm SU(3)_c}^{\rm SUSY}}{\cos \Theta(\phi_\pm)}&\lesssim& \frac{12\;\epsilon^3 \MSUSY^3 \mu_T}{(8\pi^2)^4} \left(\frac{2\pi}{\alpha_3(\MSUSY)}\right)^{6} \det(Y_u Y_d)_6 \; e^{-\frac{2\pi}{\alpha_3(\MSUSY)}} \lesssim (10^{-5}\;{\rm GeV})^4 \label{eq:SU3SUSYest} \, , \\
\frac{V_{\rm SU(2)_L}^{\rm SUSY}}{\cos \Theta(\phi_\pm)}&\lesssim& \frac{81\; \epsilon^3 \MSUSY^3 \mu_H}{\pi^2(8\pi^2)^3} \left(\frac{2\pi}{\alpha_2(\MSUSY)}\right)^{4} \det(Y_{uT} Y_{dT})_6 \; e^{-\frac{2\pi}{\alpha_2(\MSUSY)}} \lesssim (10^{-19}\;{\rm GeV})^4 \, .  \label{eq:SU2SUSYest}
\ee

The upper bound means that we are choosing the universe that gives the largest contribution between the four possibilities in Fig.~\ref{fig:uni1} and~\ref{fig:uni2}. In practice we are taking $\mu_{H, T}=\MGUT$, assuming that the triplets are at $\MSUSY$ or below in the $SU(2)_L$ calculation, and that either the triplets or the doublets are at $\MSUSY$ or below in the $SU(3)_c$ calculation. 

Note that there are additional diagrams for SUSY $SU(5)$, $SU(3)_c$ and $SU(2)_L$ where higgsino or tripletino legs and gaugino legs are closed together, but they give contributions suppressed by $B\mu/(\mu_{H,T} \MSUSY) \simeq 10^{-10}\epsilon$, compared to Eqs.~\eqref{eq:SU5n}, \eqref{eq:SU3SUSYest}, and~\eqref{eq:SU2SUSYest}, which are already negligible on the scale of $\Lambda_{\rm SM}^4$. A third class of diagrams, where gaugino and quark legs are closed together and coupled to a pair of quark and tripletino legs via a squark loop also exists and it is suppressed by $\tilde M_3/\mu_{H,T} \simeq 10^{-10}$, compared to Eqs.~\eqref{eq:SU5n}, \eqref{eq:SU3SUSYest}, and~\eqref{eq:SU2SUSYest}. We discuss these contributions more carefully in Section~\ref{sec:SR}.

\paragraph{SM Instantons when $\mathbf{\langle T \rangle =0}$} Below the SUSY breaking scale $\MSUSY$ we can compute instanton contributions to the potential within the SM. We recall that only one doublet (or one triplet) can exist parametrically below $\MSUSY$ and we do the calculation including a single doublet $H$ or triplet $T$. We assume that we are in the alignment limit of the 2HDM for both $H$ and $T$, so that they have SM-like couplings to fermions. 

\begin{figure}
     \centering
     \begin{subfigure}[b]{0.32\textwidth}
         \centering
         \includegraphics[width=\textwidth]{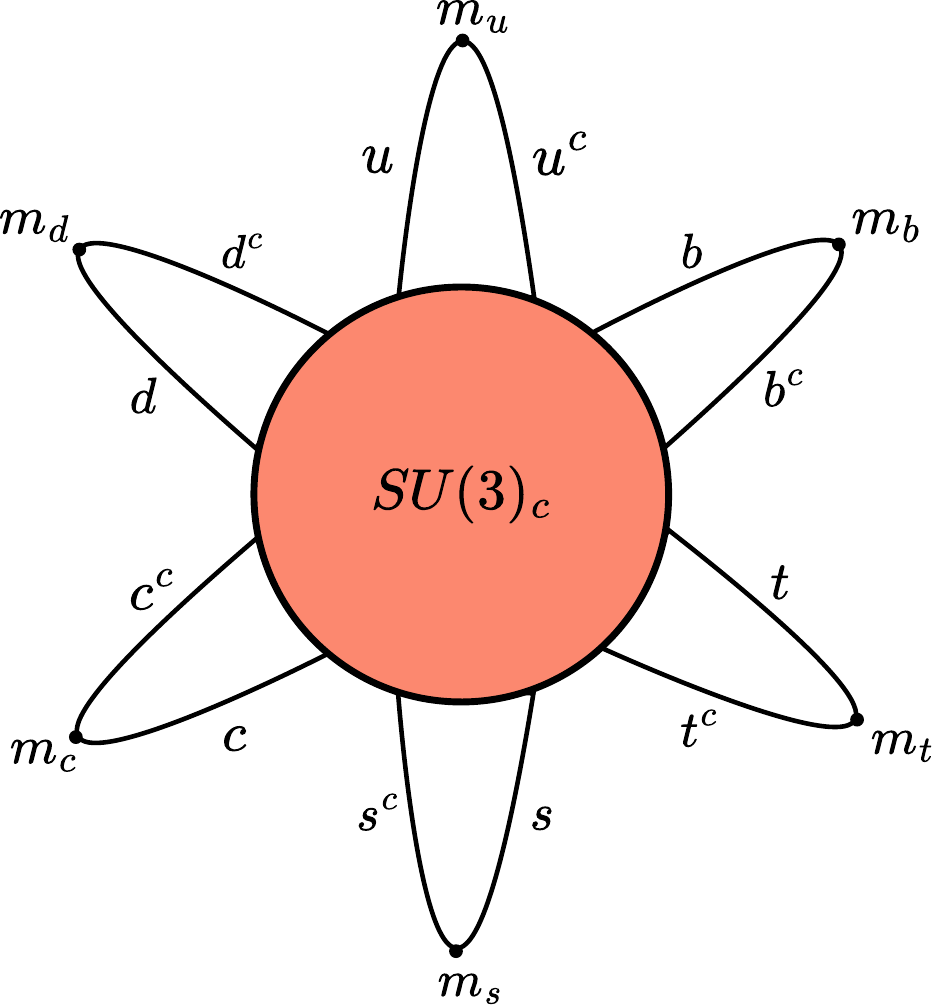}
         \caption{$SU(3)_c$ Mass insertions}
         \label{SU(3)cmass}
     \end{subfigure}
     \hfill
     \begin{subfigure}[b]{0.32\textwidth}
         \centering
         \includegraphics[width=\textwidth]{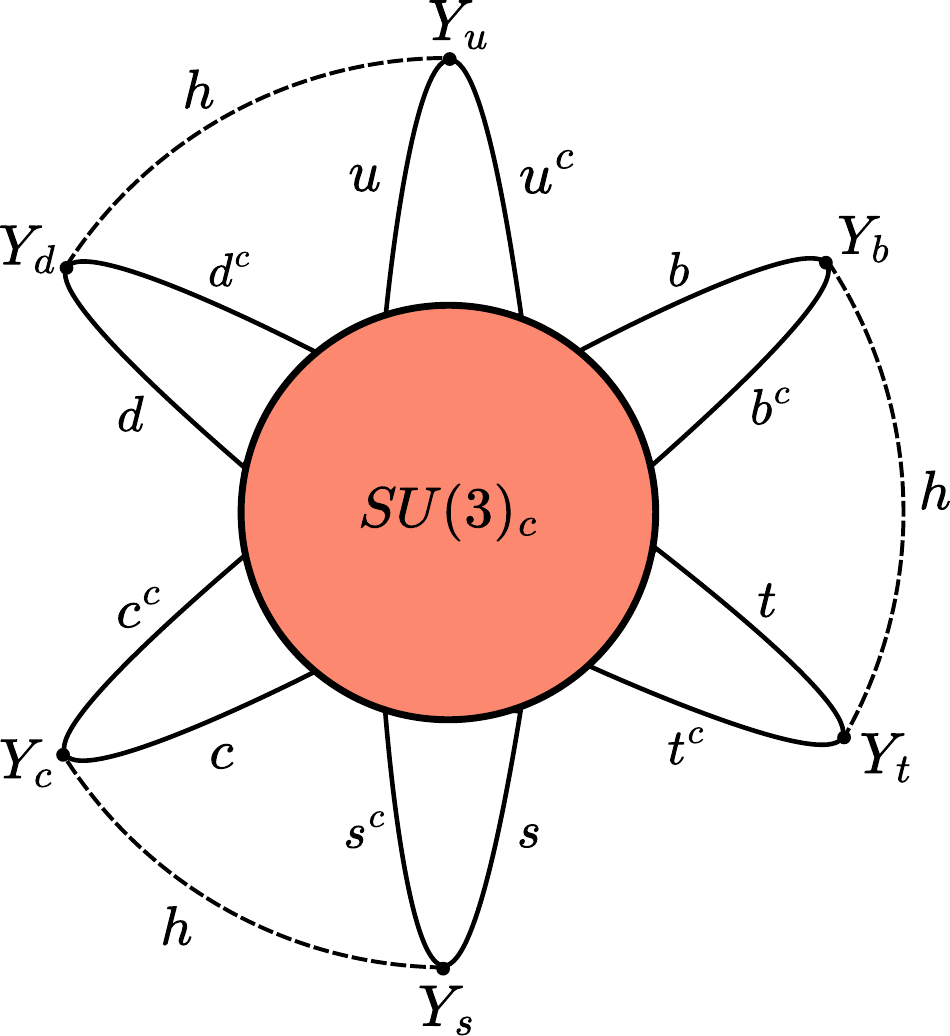}
         \caption{$SU(3)_c$ Doublet loops}
         \label{SU(3)chiggs}
     \end{subfigure}
     \hfill
     \begin{subfigure}[b]{0.32\textwidth}
         \centering
         \includegraphics[width=\textwidth]{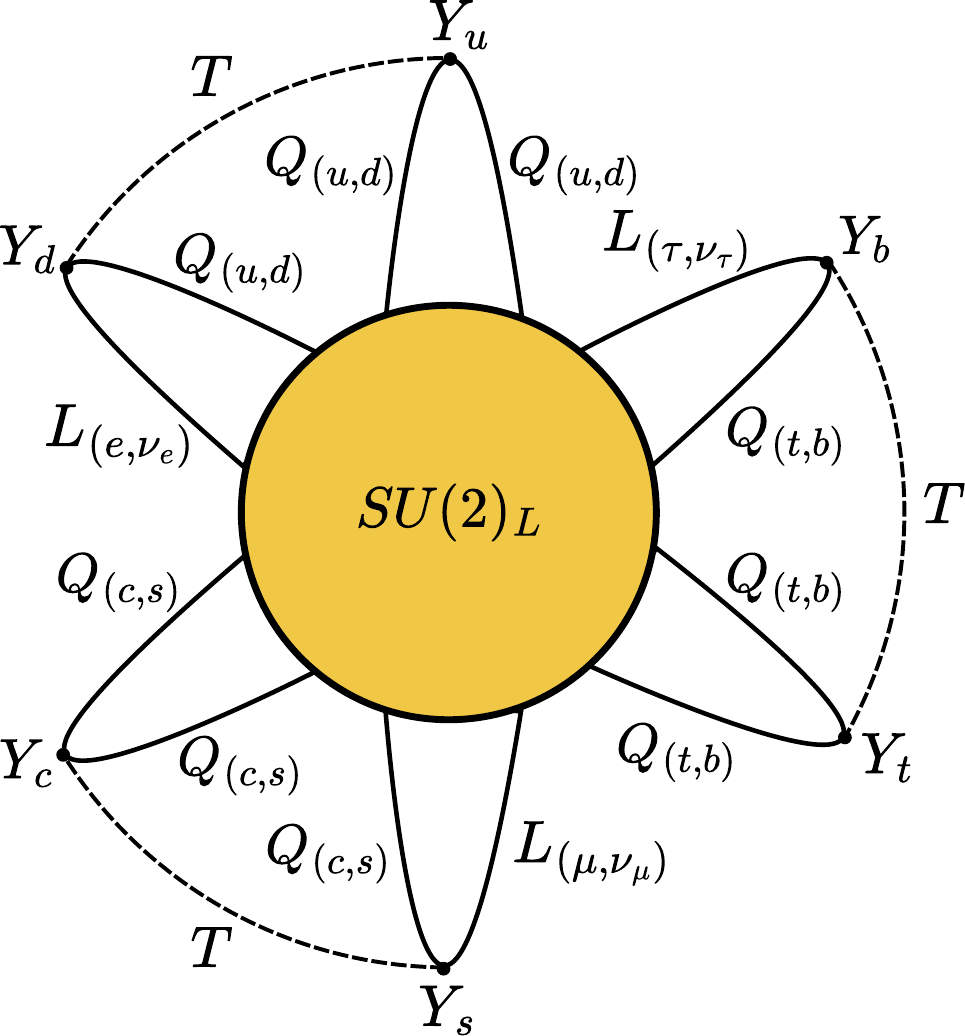}
         \caption{$SU(2)_L$ Triplet loops}
         \label{SU(2)Ltriplet}
     \end{subfigure}
        \caption{QCD and $SU(2)_L$ instanton contributions to the axion-like $\phi_\pm$ potentials in universes where $\langle T \rangle =0$. The diagrams are shown for instanton sizes $\rho^{-1} > m_t$, at lower energies heavy quarks must be integrated out. In the main text we give a general expression valid for any number of light quarks. The $SU(3)_c$ 't Hooft vertex can be closed also with Triplet loops, we discuss this contribution in the main text.}
        \label{fig:SMinst}
\end{figure}

The Yukawa couplings of the triplet provide the explicit violation of $B+L$ needed to generate a potential also from $SU(2)_L$ instantons. They contribute with a single diagram depicted in the right panel of Fig.~\ref{fig:SMinst} 
\be
\frac{V_{\rm SU(2)_L}}{2\cos \Theta(\phi_\pm)}&=&\det(Y_{uT} Y_{dT})_{2n_g} 3^{n_g}(n_g!)e^{-\alpha(1)+(4n_g-1)\alpha(1/2)}\nonumber\\
&\times &\int_{\MSUSY^{-1}}^\infty\frac{d\rho}{\rho^5}d_2(\rho)e^{-2\pi^2 \rho^2 \max[\langle H \rangle, f_\pi]^2}\rho^{2n_g}\left[ \frac{4\rho^4}{\pi^4}\int_{\{x_i\}}\frac{D_{T}(x_1-x_2)}{(x_1^2+\rho^2)^3(x_2^2+\rho^2)^3} \right]^{n_g} \, , \nn \\\label{eq:SU2inst}
\ee
where $\alpha(1)\simeq 0.44$ and $\alpha(1/2)\simeq 0.15$. The term $e^{-2\pi^2 \rho^2 \max[\langle H \rangle, f_\pi]^2}$ is due to the breaking of $SU(2)_L$ by the doublets VEVs or by QCD~\cite{Affleck:1980mp, Csaki:1998vv, Nielsen:1999vq, Csaki:2019vte} and regulates the low energy behavior of the integral. When the triplets do not have a VEV, the $SU(3)_c$ contributions come from three different diagrams, two of which are depicted in Fig.~\ref{fig:SMinst}. The two diagrams give
\be
\textnormal{Quark masses:}\quad  \frac{V_{\rm SU(3)_c}^{(1)}}{2\cos \Theta(\phi_\pm)}&=& e^{-\alpha(1)+(4n_g-n_T-2)\alpha(1/2)} \int\frac{d\rho}{\rho^5}d_3(\rho)\prod_{f=1}^{2n_g}\left(\rho m_f\right)\, , \label{eq:QCDUVMasses}\\
\textnormal{Higgs loops:}\quad  \frac{V_{\rm SU(3)_c}^{(2)}}{2\cos \Theta(\phi_\pm)}&=& e^{-\alpha(1)+(4n_g-n_T-2)\alpha(1/2)}2^{n_g}(n_g!)\textnormal{det}(Y_u Y_d)_{2n_g}\nn \\
&\times& \int  \frac{d\rho}{\rho^{5}} d_3(\rho) \rho^{2n_g}\left[\frac{4\rho^4}{\pi^4}\int_{\{x_i\}}\frac{D_{H}(x_1-x_2)}{(x_1^2+\rho^2)^3(x_2^2+\rho^2)^3} \right]^{n_g}\, .
\label{eq:QCDUVHiggs}
\ee
where $n_T$ is the number of color tirplets below $\widetilde{M}_S$. As in the supersymmetric case, there is an extra diagram where quark lines are closed by triplet loops. One can obtain its contribution by dividing the Higgs loops contribution by $2^{n_g}$ and replacing the doublet propagator with a triplet propagator, plus the same replacement for the Yukawa couplings discussed for Eq.~\eqref{eq:SU3SUSYest}. Note that only the quarks with mass $< \rho^{-1}$ contribute to the integral, so depending on the scale we have to evaluate the above expressions at different values of $n_g$.

The $SU(3)_c$ integrals in Eqs.~\eqref{eq:QCDUVMasses} and~\eqref{eq:QCDUVHiggs} are IR dominated, but when the gauge coupling becomes non-perturbative, the dilute instanton gas approximation that we have used to compute the potential breaks down. Additionally, one cannot prove anymore that instantons are the dominant saddle of the path integral. Analytical arguments at large-$N$ show that instantons cannot reproduce the main qualitative features of the axion potential from the strong dynamics~\cite{Witten:1978bc, DiVecchia:1980yfw, Kawarabayashi:1980dp, Ohta:1981ai, Csaki:2023yas}. Therefore we use these integrals only to compute the UV contribution to the potential. For Eq.~\eqref{eq:QCDUVHiggs} this contribution is parametrically similar to Eq.~\eqref{eq:SU3SUSYest}, but smaller, since we have to replace $\mu_T$, that we took at $\MGUT$ when evaluating Eq.~\eqref{eq:SU3SUSYest}, with $\MSUSY=10^6$~GeV. Eq.~\eqref{eq:QCDUVMasses} gives a much smaller contribution, suppressed by $\prod_{f =1}^{2n_g} (m_f/\MSUSY)$ compared to Eq.~\eqref{eq:QCDUVHiggs}. 

The $SU(2)_L$ integral in Eq.~\eqref{eq:SU2inst} is cutoff in the IR by EW symmetry breaking, and we do not need to regulate it. We find that the integral is dominated by contributions at $\MSUSY$. If the triplets are light ($m_{T_{u,d}}\lesssim \MSUSY$), these contributions can be obtained from Eq.~\eqref{eq:SU2SUSYest} with the replacements $\epsilon \to 1$ and $\mu_H\to\MSUSY$, which makes them much smaller than the upper bound in Eq.~\eqref{eq:SU2SUSYest}.

\paragraph{SM Instantons for the  $\mathbf{\langle T \rangle >0}$ case} If the triplet has a VEV, at energies below $\sim g_3 \langle T \rangle$ we have to include instanton contributions from $SU(2)_c$ that come from the same diagrams that we discussed for $SU(3)_c$, with the difference that quark masses come from the triplet vev $\langle T \rangle$ and are not necessarily in the form $q q^c$. This does not change the parametrics of the diagrams, as the selection rules of $U(1)_A$ still enforce the presence of a factor $\det(Y_{uT} Y_{dT})$ in front of the result which is numerically close to $\det(Y_u Y_d)$. 
As for ordinary QCD we use these instanton integrals to estimate the UV contribution to the $\phi_\pm$ potentials and discard the IR part where $SU(2)_c$ confines and instanton contributions are not well-defined. Parametrically we get
\be
\frac{V_{\rm SU(2)_c}}{2\cos \Theta(\phi_\pm)} \simeq M_g^4\left(\frac{2\pi}{\alpha_3(M_g)}\right)^{4} \det(Y_{uT} Y_{dT}) \; e^{-\frac{2\pi}{\alpha_3(M_g)}} \lesssim (10^{-6}\; {\rm GeV})^4 \, ,
\ee
where $M_g \simeq g_3 \langle T \rangle$ is the gluon mass. In our Figures we include the full expression of the instanton contribution.

\subsubsection{Selection Rules and Summary of Results}\label{sec:SR}

So far we have enumerated the dominant contributions to the $\Theta(\phi_\pm)$ potentials coming from instantons. In doing so, we mainly focused on the parametrically larger terms  and only talked briefly about certain subleading contributions.  However, it is easy to derive the general form of the result using the selection rules of the anomalous $U(1)$ symmetries of the model. We illustrate this for the case of  $SU(5)$ instantons. The other cases follow the same logic and we do not present them here.

\begin{table}[!t]
\begin{center}
\begin{tabular}{|c|c|}
\hline
& $U(1)_{R}$ charges \\
\hline
$H_5$ & $R_{H_5}$ \\
\hline
$H_{\bar 5}$ &  $R_{\overline H_5}$  \\
\hline
V & 0 \\
\hline
$\Phi_{10}$ & $R_{10}$ \\
\hline
$\Phi_{\bar 5}$ & $R_{5}$ \\
\hline
$B\mu$ & $-(R_{H_5}+R_{\overline H_5})$  \\
\hline
$\widetilde m_{\lambda}$ & $-2$ \\
\hline
$\mu_5$ & $-(R_{H_5}+R_{\overline H_5}-2)$ \\
\hline
$Y_{\bar 5}$ & $-(R_5+R_{10}+R_{\overline H_5}-2)$ \\
\hline
$Y_{10}$ & $-(2R_{10}+R_{H_5}-2)$ \\
\hline
\end{tabular}
\end{center}
\caption{$R$-charge assignments in $SU(5)$ for the generalized $R$-symmetry that fixes the parametric dependences in the $\Theta(\phi_\pm)$ instanton potential. $V$ is the gauge supermultiplet.}
\label{tab:charges}
\end{table}%

In Table~\ref{tab:charges} we list the charge assignment for a generalized $R$-symmetry that fixes the dependence on supersymmetric and SUSY breaking parameters (for definitions see Eqs.~\eqref{eq:SUSYSU5}, \eqref{eq:MSSM}). This symmetry is explicitly broken by the superpotential, but the  resulting selection rules are more powerful than those of the ordinary $R$-symmetry. We also make use of the usual SM $U(1)_A$ that acts independently on each generation of fermions in the $10$ and in the $\bar 5$ matter fields. The axion-like potential from instantons looks like
\be
V_{\rm inst} = e^{i\Theta(\phi_\pm)}\Lambda_{\rm inst}^4 +{\rm h.c.}\, .
\ee
We can formally restore the anomalous symmetries of this potential by shifting $\Theta$,
\be
U(1)_{R}\; : \;\Theta &\to& \Theta - (8-4 n_g+3n_g R_{10}+n_g R_5+R_{H_5}+R_{\overline H_5}) \alpha\, , \\
U(1)_{A}\; : \;\Theta &\to& \Theta +\alpha_{5_1}+\alpha_{5_2}+\alpha_{5_3}+3\alpha_{10_1}+3\alpha_{10_2}+3\alpha_{10_3} \,  ,
\ee
where $\alpha$ is the parameter of the $U(1)_R$ transformation, and $\alpha_{5_i},\alpha_{10_i}$ of the $U(1)_A$ transformations, while $R$ are the $R$-charges in Table~\ref{tab:charges}.

These spurion transformations constrain the parametric dependence of $\Lambda^4_{\rm inst}$ on the SUSY breaking scale and Yukawa couplings. The $U(1)_A$'s together with flavor invariance require 
\be
\Lambda_{\rm inst}^4 \propto \det(Y_5 Y_{10})\, .
\ee
The $U(1)_{R}$ symmetry constrains the powers of SUSY-breaking parameters that can appear in the potential, leaving two possible terms
\be
\Lambda_{\rm inst}^4 \propto \det(Y_5 Y_{10}) \times \left\{\begin{array}{c}(\mu^*_5)(B\mu)^{n_g} (\widetilde M_\lambda^*)^{5} \\ (B\mu)^{n_g-1} (\widetilde M_\lambda^*)^{4} \end{array}\right. \, . \label{eq:sel}
\ee
This result assumes that there can't be any T SUSY-breaking terms appearing in the denominator, since the supersymmetric limit should yield $\Lambda_{\rm inst} \to 0$~\cite{Dine:1986bg}. Additionally there can also be no $\mu_5$ in the denominator, because the massless Higgsino limit should be finite. Even with these assumptions the selection rules of $U(1)_R$ cannot fully determine the structure of $\Lambda_{\rm inst}^4$, as we can always multiply the above terms by arbitrary powers of $|B\mu|$ or $|\mu_5|$ or any other parameter that transforms trivially. However they allow us to conclude that our result should be proportional to one of the two terms in Eq.~\eqref{eq:sel}. We use this as a sanity check of our explicit calculation.

The first term in Eq.~\eqref{eq:sel} is the one appearing in Eq.~\eqref{eq:SU5n}, the second one is suppressed by $\MSUSY/\MGUT$ and we have only mentioned it in the text, without giving an explicit expression. In our full calculation, we find a third parametric contribution that we mentioned below Eq.~\eqref{eq:SU2SUSYest}, which is given by the second term in Eq.~\eqref{eq:sel} times $|B\mu|$, and is even more suppressed than the two in Eq.~\eqref{eq:sel}.

\begin{figure}[!t]
\begin{center}
\includegraphics[width=0.9\textwidth]{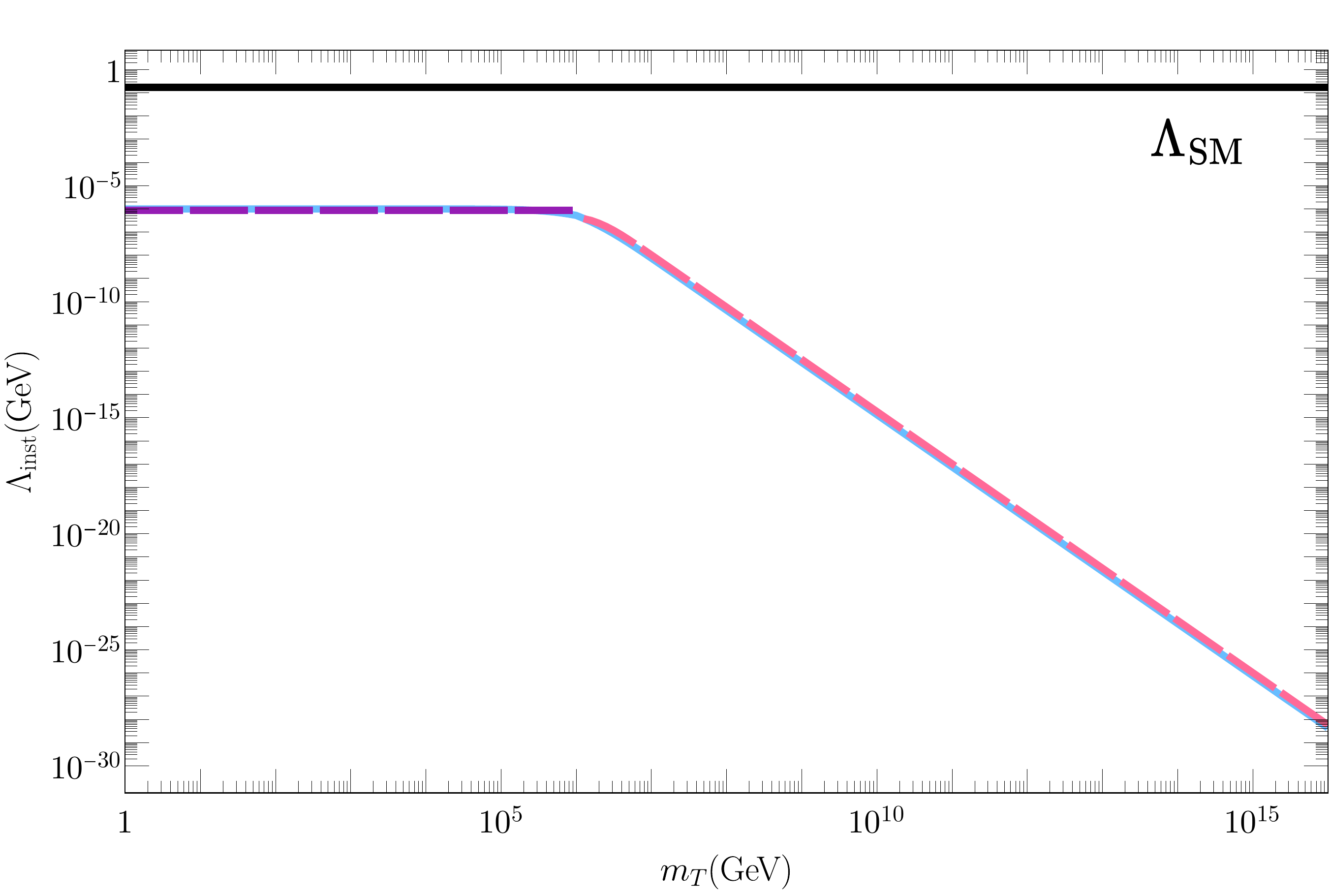}
\caption{Size of the largest instanton contribution to $V_{\phi H}$ in our Multiverse, given by SUSY QCD in Eq.~\eqref{eq:SU3SUSY}. The cerulean line is the full numerical result, the two dashed lines are analytical approximations described in Appendix~\ref{app:integrals}.}
\label{fig:inst}
\end{center}
\end{figure}

To conclude this Section, we show in Fig.~\ref{fig:inst} the size of the largest instanton contribution to $V_{\phi H}$ in our Multiverse. It is given by SUSY QCD in Eq.~\eqref{eq:SU3SUSY}. 
The solid light blue line in Fig.~\ref{fig:inst} is the full numerical result, while the two dashed lines are the analytical approximations described in Appendix~\ref{app:integrals}. When $m_T < \MSUSY$ the integral over instanton sizes is dominated at $\MSUSY$ and we can neglect the triplet masses in the propagators, obtaining Eq.~\eqref{eq:SU3SUSYest},
\be
\left(\frac{V_{\rm SU(3)_c}^{\rm SUSY}}{\cos \Theta(\phi_\pm)}\right)_{\rm light} \simeq (10^{-5}\;{\rm GeV})^4\, ,
\ee
which is represented in the plot by the purple dashed line. When $m_T \gg \MSUSY$ we integrate out the triplets as discussed in Appendix~\ref{app:integrals} and get an extra suppression
\be
\frac{V_{\rm SU(3)_c}^{\rm SUSY}}{\cos \Theta(\phi_\pm)}&\simeq& \left(\frac{V_{\rm SU(3)_c}^{\rm SUSY}}{\cos \Theta(\phi_\pm)}\right)_{\rm light} \left(\frac{\MSUSY}{m_T}\right)^{6+b_0}\, ,
\ee
where $b_0=3$ for SQCD, which reproduces the $\Lambda_{\rm inst}\sim m_T^{-\frac{9}{4}}$ scaling of the numerical result.

The main message of Fig.~\ref{fig:inst} is that instanton contributions to $V_{\phi H}$ are negligible in our Multiverse ($\Lambda_{\rm inst}\ll \Lambda_{\rm SM}$) and cannot prevent a universe from crunching.


\section{Conclusion}\label{sec:conclusion}
In this work we have presented a joint solution to the doublet-triplet splitting problem of GUTs, the electroweak hierarchy problem and the strong CP problem. Our solution relies on the existence of two light and weakly-coupled axion-like particles (ALPs) with distinct phenomenology. As shown in Eq.~\eqref{eq:mphi}, the heavier of the two ALPs ($\phi_-$) lives slightly below the QCD line in the plane of axion coupling vs mass, i.e. it is heavier than a QCD axion with the same coupling. The lighter ALP ($\phi_+$) must have a mass smaller than Hubble at the QCD phase transition in our universe, and lives on the QCD line. 

As discussed in~\cite{TitoDAgnolo:2021nhd}, $\phi_+$ can also explain all the observed dark matter energy density in our universe. If this is the case, the dark matter mass must be $\lesssim 10^{-19}$~eV and is related to the value of the neutron EDM as in Fig.~3 of~\cite{TitoDAgnolo:2021nhd}. We did not review these predictions in detail in this work, because they are the same as in~\cite{TitoDAgnolo:2021nhd}. The only phenomenological difference between our work and~\cite{TitoDAgnolo:2021nhd} is that our axion-like particles couple to the full $SU(5)$ field strengths $F_5 \widetilde F_5$ rather than just to QCD. However, this only generates a tiny contribution to their potential from $SU(2)_L$ that is completely negligible compared to the QCD contribution already present in~\cite{TitoDAgnolo:2021nhd}.

Our two ALPs can make all universes that are qualitatively different from our own crunch (i.e. quickly end up with a large negative cosmological constant) shortly after the QCD phase transition. To show this, we have computed the contributions from the IR strong dynamics (Section~\ref{sec:QCD}) and UV instantons (Section~\ref{sec:inst}) to the $\phi_\pm$ potentials in GUTs with different choices of doublets and triplets masses. We have found that instantons are negligible in most universes with $\aGUT$ close to its value in the MSSM and that QCD dominates the $\phi_\pm$ potentials. The QCD confinement scale depends on the doublet and triplet masses and VEVs, and in most universes it is much smaller than in our own. A small QCD scale results in a contribution to the $\phi_+$ potential which is too small to stabilize it and these universes crunch, as explained in Sections~\ref{sec:basic} and~\ref{sec:sliding}. $\phi_-$ crunches universes with a large Higgs VEV, where $\LQCD$ is larger than in our own.

Our calculations were performed in a supersymmetric model of unification with a fixed scale of SUSY breaking, $\MSUSY=10^6$~GeV. This choice shows that the same dynamics that solves the doublet-triplet splitting problem also solves the hierarchy problem without relying on SUSY. Qualitatively, our results apply also to non supersymmetric models of unification and we believe that the same mechanism can solve the doublet-triplet splitting problem, the electroweak hierarchy problem and the strong CP problem, also in this context. We leave the detailed description and investigation of these non-supersymmetric models to future work. 

In summary, this work shows that while naturalness can be elusive, there may still be some detectable consequences, even if the tuning that we want to explain is scanned in the Multiverse.

\appendix
\section{Instanton Integrals}\label{app:integrals}
In Section~\ref{sec:inst} we often quote approximate expressions for the instanton contributions to our axion-like potentials. We obtained them following the simple analytical approximations in this appendix.

If we close the legs of the 't Hooft operator with fermion mass insertions, we encounter integrals of the form
\be
I_m(F)&=&\int_{\MGUT^{-1}}^{\LQCD^{-1}}\frac{d\rho}{\rho^5}\left(\frac{2\pi}{\alpha_s(1/\rho)}\right)^6 e^{-\frac{2\pi}{\alpha_s(1/\rho)}}\prod_{f=1}^F\left(\rho m_f\right)\nn \\
&=& \mu^{b_0}e^{-\frac{2\pi}{\alpha_s(\mu)}}\int_{\MGUT^{-1}}^{\LQCD^{-1}}\frac{d\rho}{\rho^{5-b_0-F}}\left(\frac{2\pi}{\alpha_s(1/\rho)}\right)^6 \det m_q \nn \\
&\simeq& \frac{1}{F+b_0(\MGUT)-4}e^{-\frac{2\pi}{\aGUT}}\MGUT^{4-F}\left(\frac{2\pi}{\aGUT}\right)^6 \det m_q \nn \\
&+& \frac{1}{F+b_0(\LQCD)-4}\LQCD^{4-F}\det m_q\, ,
\ee
where we have ignored the logarithmic dependence of $\alpha_s$ on $\rho$ when not exponentiated. In the second equality we have ignored the change of $b_0$ between $\LQCD$ and $\MGUT$. In principle we should break the integral into intervals of constant $b_0$ and evaluate every single contribution. Since the procedure is straightforward we do not write it explicitly here, especially since the running will make either the deep UV or the deep IR contributions dominant (i.e. one of the two that we have written down in the third equality).

When closing fermion legs with Yukawa couplings and scalar loops we encounter integrals of the type
\be
I_H(m_H, \rho)=\frac{4\rho^4}{\pi^4}\int d^4 x_1 \int d^4 x_2 \int \frac{d^4 p}{(2\pi)^4}\frac{e^{i p\cdot (x_1-x_2)}}{(p^2+m_H^2+i \epsilon)^2}\frac{1}{(x_1^2+\rho^2)^3} \frac{1}{(x_2^2+\rho^2)^3}\, .
\ee
We can first use the identity
\be
\int d^4 x \frac{e^{-ipx}}{(x^2+\rho^2)^3}=\frac{\pi^2}{2\rho^2}(p\rho)K_1(p\rho)
\ee
that gives
\be
I_H(m_H, \rho)&=& \int\frac{d^4 p}{(2\pi)^4}\frac{[(p\rho)K_1(p\rho)]^2}{(p^2+m_H^2+i \epsilon)^2} 
=\frac{1}{8\pi^2}\int_0^\infty dy \frac{y^5 K_1(y)^2}{(y^2+m_H^2\rho^2+i\epsilon)^2} \, .
\ee
Then we first expand for small $m_H \rho$ to obtain the result valid when the scalars are light compared to the size of the instanton
\be
I^{\rm light}_H(m_H,\rho) =\frac{1}{8\pi^2}\log\left[\frac{2e^{-\left(1+\gamma_E\right)}}{ m_H \rho}\right]+\mathcal{O}( m_H^2\rho^2)\, .
\ee
Expanding for large $m_H\rho$ we get instead 
\be
I^{\rm heavy}_H(m_H,\rho) &\simeq&  \frac{1}{8\pi^2}\frac{1}{( m_H\rho)^4}\int_0^{m_H\rho }dy~ y^5 K_1(y)^2 \nn \\
&\simeq & \frac{1}{8\pi^2}\left[\frac{8}{5}\frac{1}{( m_H \rho)^4}-\frac{\pi}{4}e^{-2 m_H\rho}\left(1+\mathcal{O}( m_H\rho)^{-1}\right)\right]\, .
\ee
In our Figures we keep more terms in the $1/\rho m_H$ expansion to improve agreement with our numerical results. The integrals over instanton sizes are saturated at $\rho^{-1} \simeq m_H$ which makes the expansion converge slowly, even if we resummed the leading terms to get the factor $e^{-2 m_H \rho}$ in the previous expression.
Finally, in the case of constrained instantons we have
\be
I_c&=& \int^{\infty}_{\MGUT^{-1}}d\rho \rho^{p-5}e^{-2\pi^2 v^2\rho^2} e^{-\frac{2\pi}{\alpha_W(\rho)}} \nn \\
&=&e^{-\frac{2\pi}{\aGUT}}\left[\frac{\MGUT^{4-p}}{4-p}+\frac{(\sqrt{2}\pi v)^{4-p}}{2}\left(\frac{\MGUT}{\sqrt{2}\pi v}\right)^{b_{0W}}\Gamma\left(\frac{p-4}{2}\right)+\mathcal{O}\left(\frac{2\pi^2 v^2}{\MGUT^2} \right)\right]\, .
\ee
where $b_{0W}$ is the $SU(2)_L$ $\beta$-function coefficient
\be
\frac{d g}{d\log\mu}= - \frac{b_{0W}}{16\pi^2}g^3(\mu)\, ,
\ee
that we assumed to be constant between $\MGUT$ and $v$, when this is not the case, the integral can be broken into intervals of constant $b_{0W}$.

\section{Pion Masses with no Vevs}\label{app:goldstones}
The computation of the axion mass in universes with no vevs differs from the standard SM computation. In this appendix we give more details compared to the main text where we only listed the relevant operators in the Chiral Lagrangian and the final result.

We take the quark fields to transform under $SU(6)_L \times SU(6)_R$ as
\be
q \to L q\, , \quad q^c \to q^c R^\dagger\, ,
\ee
so we can think of $q$ as a column vector and $q^c$ as a row vector. We choose the embedding
\be
q = \left( u \;\; c\;\; t \;\; d\;\; s \;\; b \right)^T\, .
\ee
Then we can write the pion matrix as $U \sim \langle q q^c \rangle/(\Lambda_{\rm QCD} f_\pi^2)$, which implies the transformation property
\be
U \to L U R^\dagger\, .
\ee
To write the Chiral Lagrangian without doublets or triplets vevs, it is useful to promote the Yukawa couplings to spurions. The Yukawa Lagrangian for quarks, below the GUT scale, reads
\be
\mathcal{L} \supset -\frac{1}{2}Q^T Y_{uT} Q T_u + u^c Y_{dT} (d^c)^T T_d + u^c Y_u Q H_u + d^c Y_d Q H_d + {\rm h.c.}\, , \label{eq:TYukApp}
\ee
where all the Yukawa couplings are $3\times 3$ matrices in flavor space. In most of our universes the masses of doublets and triplets are much larger than the QCD scale, so we can integrate them out before matching to the Chiral Lagrangian. Additionally, as explained in the main text, doublets and triplets cannot be simultaneously lighter than the GUT scale. Therefore we ignore the pair of scalars at the GUT scale and consider only the operators obtained integrating out the lightest scalars. If the doublets are light we have
\be
\mathcal{L}_H \supset \left(B\mu \frac{(Y_uQ u^c)(Y_d Q d^c)}{m_U^2 m_D^2}+{\rm h.c.}\right)+\frac{|Y_u Q u^c|^2}{m_U^2}+\frac{|Y_d Q d^c|^2}{m_D^2}\, . \label{eq:doublet6}
\ee
Instead light triplets give
\be
\mathcal{L}_T \supset \left(B\mu \frac{(Y_{uT} Q Q)(Y_{dT} u^c d^c)}{2m_{U_T}^2 m_{D_T}^2}+{\rm h.c.}\right)+\frac{|Y_{uT} Q Q|^2}{4m_{U_T}^2}+\frac{|Y_{dT} u^c d^c|^2}{m_{D_T}^2}\, . \label{eq:triplet6}
\ee
To make the transition to the Chiral Lagrangian simpler we can write these operators in terms of our six-flavor quark vector $q$, 
\be
\mathcal{L}_H &\supset & \left(B\mu \frac{(q^c Y_u^6 q)(q^c Y_d^6 q)}{m_U^2 m_D^2}+{\rm h.c.}\right)+\frac{|q^c Y_u^6 q|^2}{m_U^2}+\frac{|q^c Y_d^6 q|^2}{m_D^2}\, , \nn \\
\mathcal{L}_T &\supset & \left(B\mu \frac{(q^T Y_{uT}^6 q)(q^c Y_{dT}^6 (q^c)^T)}{4m_{U_T}^2 m_{D_T}^2}+{\rm h.c.}\right)+\frac{|q^T Y_{uT}^6 q|^2}{4m_{U_T}^2}+\frac{|q^c Y_{dT}^6 (q^c)^T|^2}{4m_{D_T}^2}\, .
\ee
and embed the $3\times 3$ Yukawa matrices into $6\times 6$ matrices
\be
Y_u^6=\left(\begin{array}{cc} Y_u & Y_u \\ 0 & 0\end{array}\right)\, , &\quad & Y_d^6=\left(\begin{array}{cc} 0 & 0 \\ Y_d & Y_d\end{array}\right)\, , \nn \\ Y_{uT}^6=\left(\begin{array}{cc} 0 & Y_{uT} \\ -Y_{uT} & 0\end{array}\right)\, , &\quad & Y_{dT}^6=\left(\begin{array}{cc} 0 & Y_{dT} \\ -Y_{dT} & 0\end{array}\right)\, .\label{eq:Y6}
\ee
Note that we can diagonalize all $3\times 3$ Yukawa matrices as in the SM, but in the triplet case there is no CKM, since $Y_{uT}$ must be symmetric according to Eq.~\eqref{eq:TYukApp} and we can diagonalize it with a common rotation of the two components of the $Q$ doublet. We take this into account in the above expressions (i.e we take the Yukawa matrices to be already diagonalized, $Y=Y^\dagger$).
The $6\times 6$ Yukawa matrices have simple spurion transformations
\be
Y_{uT}^6 &\to& L^* Y_{uT}^6 L^\dagger\, , \quad Y_{dT}^6 \to R Y_{dT}^6 R^T\, , \nn \\
Y_u^6 &\to& R Y_u^6 L^\dagger\, , \quad Y_d^6 \to R Y_d^6 L^\dagger\, ,
\ee
that allow us to match Eq.~\eqref{eq:doublet6} to the Chiral Lagrangian terms 
\be
\mathcal{L}_{\pi H} \supset \Lambda_{\rm QCD}^2 f_\pi^4\left(\frac{B\mu}{m_U^2 m_D^2}{\rm Tr}[Y_u^6 U Y_d^6 U]+{\rm h.c.}+\frac{|{\rm Tr}[Y_u^6 U]|^2}{m_U^2}+\frac{|{\rm Tr}[Y_d^6 U]|^2}{m_D^2}\right)\, .
\ee
Every operator comes with an unknown $\mathcal{O}(1)$ coefficient that here and in the following we set to one.
We can proceed in the same way for Eq.~\eqref{eq:triplet6} and obtain
\be
\mathcal{L}_{\pi T} \supset \Lambda_{\rm QCD}^2 f_\pi^4\left(\frac{B\mu}{4m_{U_T}^2 m_{D_T}^2}{\rm Tr}[Y_{uT}^6 U Y_{dT}^6 U^T]+{\rm h.c.}\right)\, .
\ee
The last two operators in  Eq.~\eqref{eq:triplet6} contain four left-handed quarks or four right-handed quarks and do not match to operators that can give a pion mass term relevant to our axion-like potential calculation.

As an example we show how to compute the
$\phi_\pm$ masses in the case where doublets are light and $m_U^2 \ll m_D^2$. We first write the Chiral Lagrangian
\be
\mathcal{L}_\pi&=&\frac{f_\pi^2}{4}{\rm Tr}[(\partial_\mu U)^\dagger \partial^\mu U] + \Lambda_{\rm QCD}^2 f_\pi^4\left[\frac{B\mu}{m_U^2 m_D^2} \left({\rm Tr}[Y_u^6 U Y_d^6 U]+{\rm h.c.}\right) + \frac{1}{m_U^2} |{\rm Tr}[Y_u U]|^2\right] \nn \\
&-& \frac{a}{4}\Lambda_{\rm QCD}^2 f_\pi^2 \left(- \Theta(\phi_\pm) - i \log \det U \right)^2\, , 
\ee
where in the second line we included the leading large-$N$ operator coming from the $U(1)_A$ anomaly~\cite{Witten:1980sp, Witten:1978bc, DiVecchia:1980yfw,Kawarabayashi:1980dp,Ohta:1981ai} and we defined the field $\Theta$ as in the main text,
\be
\Theta(\phi_\pm)\equiv \frac{\phi_-}{F_-}+\frac{\phi_+}{F_+}+\overline{\theta}\, .
\ee
To compute the mass of $\Theta$ we first expand the pions' potential around its minimum. The symmetries of the problem forbid electromagnetically charged pions from getting a VEV, and therefore we can have VEVs only on the diagonal of the pion matrix. Additionally, the VEV does not need to respect the $SU(6)_L\times SU(6)_R \to SU(6)_V$ breaking pattern from the strong dynamics, since the Yukawa couplings break explicitly these symmetries and generate a different breaking pattern. Therefore the most general VEV for the pion matrix must be in the form 
\be
\langle U_{ij} \rangle = e^{i \phi_i} \delta_{ij}\, , \label{eq:PVEV}
\ee
i.e. it is not proportional to the identity as it would be for $SU(6)_L\times SU(6)_R \to SU(6)_V$, but it can have VEVs only on the diagonal. From Eq.~\eqref{eq:PVEV} we can compute the potential at the minimum
\be
V(\phi_i) &=& \frac{a}{4}\Lambda_{\rm QCD}^2 f_\pi^2 \left(\langle \Theta(\phi_\pm) \rangle - \sum_{i=1}^6 \phi_i\right)^2 \nn \\
&-& \frac{\Lambda_{\rm QCD}^2 f_\pi^4 B\mu}{m_U^2 m_D^2}  \left[y_u y_d \cos(\phi_1 + \phi_4) + y_c y_s \cos(\phi_2+\phi_5) + y_b y_t \cos(\phi_3+\phi_6)\right] \nn \\
&-&  \frac{\Lambda_{\rm QCD}^2 f_\pi^4}{m_U^2}  \left[y_c y_u \cos(\phi_1-\phi_2)+ y_t y_u \cos(\phi_1-\phi_3)+ y_c y_t \cos(\phi_2 - \phi_3) \right]\, . \label{eq:Vphi}
\ee
The minimum of~\eqref{eq:Vphi} is then given by
 \be
&& \langle \Theta(\phi_\pm) \rangle - \sum_{i=1}^6 \phi_i = 0\, , \nn \\
&&\phi_1 = \phi_2 = \phi_3 = - \phi_4 = - \phi_5 = - \phi_6 \, ,\label{eq:min}
\ee
modulo redundancies from the periodicity of the cosines.
Our axion-like particles get a small VEV from the QCD-independent part of their potential, but we neglect it in the above equations and set $\langle \Theta(\phi_\pm) \rangle = \sum_{i=1}^6 \phi_i$.

We can see from Eq.~\eqref{eq:min} that our potential has flat directions, but we assume that they are lifted by higher-order effects in the Yukawas and choose the CP-conserving minimum $\phi_i = 0$.
We can now expand around the minimum and compute the quadratic part of the potential
\be
V_2 &=&  \frac{M_{ij}^2}{2} \pi^i \pi^j + \frac{a}{4} \Lambda_{\rm QCD}^2 \left(f_\pi \Theta-2\sqrt{3} \pi^6\right)^2\, , \\
M_{ij}^2&\equiv& \frac{\Lambda_{\rm QCD}^2 f_\pi^2 B\mu}{m_U^2 m_D^2}\left( \frac{1}{2}{\rm Tr}[\{Y_u^6,Y_d^6\} \{T^i, T^j\}] + {\rm Tr}[Y_u^6 T^i Y_d^6 T^j+(i\leftrightarrow j)] + {\rm h.c.}\right) \nn \\
&+& \frac{\Lambda_{\rm QCD}^2 f_\pi^2}{m_U^2}\left\{\frac{1}{2}{\rm Tr}[Y_u^6 T^i T^j]{\rm Tr}[(Y_u^6)^\dagger]+\frac{1}{2}{\rm Tr}[(Y_u^6)^\dagger T^i T^j]{\rm Tr}[Y_u^6] \right. \nn \\
&-& \left.{\rm Tr}[Y_u^6 T^i ]{\rm Tr}[(Y_u^6)^\dagger T^j]+ (i\leftrightarrow j)\right\}\, . 
\ee
To get the mass of $\Theta$ we just integrate out the pions at the quadratic level. We can focus just on the Cartan subalgebra of $SU(6)$, as only neutral pions can mix with $\Theta$. The only subtlety is that $SU(2)_L$ is unbroken and we have to make a gauge transformation that removes the corresponding pions from $M_{ij}^2$ before inverting it. Then we are left with a simple set of linear equations for the $\pi$'s as a function of $\Theta$ that we can solve to get Eq.~\eqref{eq:lightHnoVEV} in the main text.

\acknowledgments

We would like to thank Max Ruhdorfer and Emile Pangburn for many useful discussions. CC is supported in part by the NSF grant PHY-2309456. EK is supported in part by NSF-BSF grant 2022713. CC and EK are supported in part by the US-Israeli BSF grant 2016153. RTD and PS acknowledge ANR grant ANR-23-CE31-0024 EUHiggs for partial support.

\bibliography{refs}
\bibliographystyle{JHEP}
\end{document}